\documentclass[12pt]{article}
\usepackage{amsmath}
\usepackage{pstricks}
\usepackage{cite,times}
\usepackage[margin=1in,a4paper,includeheadfoot]{geometry}
\usepackage[small]{titlesec}

\numberwithin{equation}{section}
\newcommand{\rme}{\textrm{e}}
\newcommand{\rmi}{\textrm{i}}
\newcommand{\eps}{\varepsilon}
\newcommand{\wt}{\widetilde}
\newcommand{\const}{\textrm{const}\times}
\newcommand{\MP}[4]{P_{#1}^{(#2)}\!\left(#3;#4\right)}

\newcommand{\F}[4]{%
    {}\,{}_2F_1\biggl(%
    \genfrac{}{}{0pt}{}{#1\,,\:#2}{#3}
    \bigg\vert#4\biggr)}
\newcommand{\Fthreetwo}[6]{%
    \,{}_3F_2\biggl(%
    \genfrac{}{}{0pt}{}{#1\,,\:#2\,,\:#3}{#4\,,\:#5}
    \bigg\vert#6\biggr)}
\newcommand{\Ffourthree}[8]{%
    \,{}_4F_3\biggl(%
    \genfrac{}{}{0pt}{}{#1\,,\:#2\,,\:#3\,,\:#4}{#5\,,\:#6\,,\:#7}
    \bigg\vert#8\biggr)}
\renewcommand{\Re}{\mathrm{Re}\,}

\titleformat{\section}{\bf\large}{\thetitle.\ \;}{0pt}{}
\titleformat{\subsection}{\bf}{\thetitle.\ \;}{0pt}{}

\begin{document}
\begin{titlepage}
\vspace*{0in}

\begin{center}
\Large\textbf{
Square ice, alternating sign matrices,
and classical orthogonal polynomials
}
\end{center}
\bigskip

\begin{center}
\large{
F. Colomo and A. G. Pronko\footnote{On leave of absence from
St-Petersburg Department of V. A. Steklov
Mathematical Institute
of Russian Academy of Sciences,
Fontanka 27, 191023 St-Petersburg, Russia.}
}
\end{center}

\begin{center}
\textsl{
I.N.F.N., Sezione di Firenze,
and Dipartimento di Fisica, Universit\`a di Firenze,\\
Via G. Sansone 1, 50019 Sesto Fiorentino (FI), Italy
}
\end{center}

\bigskip
\centerline{\textbf{Abstract}}
\smallskip

The six-vertex model with domain wall boundary conditions, or
square ice, is considered for particular values of its parameters,
corresponding to  $1$-, $2$-, and $3$-enumerations of alternating sign
matrices (ASMs).
Using  Hankel determinant representations for the partition
function and the boundary correlator of homogeneous square ice,
it is shown how the ordinary and refined enumerations can
be derived  in a very simple and straightforward way.
The derivation is based on the standard relationship
between Hankel determinants and orthogonal polynomials.
For the particular sets of parameters corresponding
to $1$-, $2$-, and $3$-enumerations of ASMs, the Hankel determinant
can be naturally related to  Continuous Hahn, Meixner-Pollaczek,
and Continuous Dual Hahn polynomials,  respectively.
This observation allows for a unified and simplified treatment
of ASMs enumerations.  In particular, along the lines of the proposed
approach, we provide a complete solution to the long standing problem
of the refined 3-enumeration of AMSs.

\end{titlepage}
\setcounter{page}{2}
\section{Introduction}

The six-vertex model
on a square lattice with domain wall
boundary conditions (DWBC) was introduced in \cite{K-82} and
subsequently solved in \cite{I-87}, where a determinant formula for the
partition function was obtained and proven (see also \cite{ICK-92}).
Analogous determinant formulae has been given also
for the boundary one point correlation functions \cite{BPZ-02}.
The  model, in its inhomogeneous formulation, i.e., with
position-dependent Boltzmann weights,
was originally proposed within the theory of
correlation functions of quantum integrable models, in the framework
of the quantum inverse scattering method \cite{KBI-93}.
The model was later found to be deeply related with the problems
of enumeration of alternating sign matrices (ASMs)
\cite{MRR-82,MRR-83,RR-86,Z-96a,Ku-96,Br-99} and domino tilings
\cite{EKLP-92a,EKLP-92b,CEP-96,JPS-98}. It should be mentioned that ASM
enumerations  appear to be in turn deeply related with quantum
spin chains and some loop models, via Razumov-Stroganov
conjecture  \cite{RS-01}; for recent works, see for instance
Ref.~\cite{DfZj-04} and references therein.

In its homogeneous version, the six-vertex model with DWBC
admits usual interpretation as a model of statistical mechanics
with fixed boundary conditions,
and it may be seen as a variation of the original model
with periodic boundary conditions \cite{L-67,L-67a,S-67,LW-72}.
The latter was originally proposed as a model for
two-dimensional ice (hence the alternative  denomination: `square ice'),
and has been for decades a paradigmatic
one in statistical mechanics \cite{B-82}.
Till now, specific results for the six-vertex model with DWBC
at particular values of its parameters (in fact, mainly derived
within the context of ASMs)
were obtained from general results for the inhomogeneous version,
first specializing the parameters to the considered case,
and then performing the homogeneous limit at the very end, and hence
once for  each particular case. Each time, the homogeneous
limit was an hard task on its own right,
and a specific approach was devised  to work it out in each
single case \cite{Ku-96,Z-96b,S-02}.

The purpose of the present paper  is to explain how one can proceed
the other way around,  first performing the homogeneous limit
once for all for the model with generic vertex weights,
and then specializing  the result to the case of interest.
The homogeneous limit has already been done  in \cite{ICK-92}
for the partition function,
and in \cite{BPZ-02} for the boundary one point correlation functions,
resulting  in Hankel determinant representations for these quantities.
Here we show  how, by specializing  the parameters to some particular values,
and exploiting the standard relationship
between Hankel determinants and orthogonal polynomials,
all previously known results concerning ordinary and
refined enumerations can be derived  in a very simple  and straightforward
way.  It appears that for the particular sets of parameters corresponding
to $1$-, $2$-, and $3$-enumerations of ASMs, the Hankel determinant
is naturally related to  Continuous Hahn, Meixner-Pollaczek,
and Continuous Dual Hahn polynomials,  respectively.
The approach which we propose here
allows for a unified and simplified treatment
of ASMs enumerations, including the problem of refined $3$-enumeration.

The paper is organized as follows. In the next Section we
recall  the definition of the model  and some related results,
with particular attention
to the Hankel determinant representations for the partition function
and the one-point boundary correlator.
In Section 3, known results on ASMs are
reviewed, together with their connection with the square ice.
In Section 4, we show how the Hankel determinant entering the partition
function of  square ice can be reinterpreted, in the three cases
corresponding to  $1$-, $2$- and $3$-enumeration of ASMs,
as the Gram determinant for an appropriate  choice of
Continuous Hahn, Meixner-Pollaczek, and Continuous Dual Hahn
polynomials, respectively, thus providing
a very simple and straightforward derivation of the known results
for ASM enumerations.
In Section 5 we exploit the fact that these polynomials,
being of hypergeometric type, satisfy some finite difference equations,
which can be translated  into recurrence relations for the
one-point boundary correlator. These recurrences in turn
can be solved, thus giving the results for the refined
$1$-, $2$- and, especially interesting, $3$-enumeration of ASMs.
We conclude in Section 6 with a  discussion of the proposed approach.

\section{Square ice with DWBC}\label{sec.6vm}

\begin{figure}[b]
\unitlength=1mm
\begin{center}
\begin{picture}(124,30)
\put(7,20){\line(0,1){10}}
\put(7,22.5){\vector(0,-1){1}}
\put(7,27.5){\vector(0,-1){1}}
\put(2,25){\line(1,0){10}}
\put(4.5,25){\vector(-1,0){1}}
\put(9.5,25){\vector(-1,0){1}}
\put(27,20){\line(0,1){10}}
\put(27,22.5){\vector(0,1){1}}
\put(27,27.5){\vector(0,1){1}}
\put(22,25){\line(1,0){10}}
\put(24.5,25){\vector(1,0){1}}
\put(29.5,25){\vector(1,0){1}}
\put(52,20){\line(0,1){10}}
\put(52,22.5){\vector(0,-1){1}}
\put(52,27.5){\vector(0,-1){1}}
\put(47,25){\line(1,0){10}}
\put(49.5,25){\vector(1,0){1}}
\put(54.5,25){\vector(1,0){1}}
\put(72,20){\line(0,1){10}}
\put(72,22.5){\vector(0,1){1}}
\put(72,27.5){\vector(0,1){1}}
\put(67,25){\line(1,0){10}}
\put(69.5,25){\vector(-1,0){1}}
\put(74.5,25){\vector(-1,0){1}}
\put(97,20){\line(0,1){10}}
\put(97,22.5){\vector(0,1){1}}
\put(97,27.5){\vector(0,-1){1}}
\put(92,25){\line(1,0){10}}
\put(94.5,25){\vector(-1,0){1}}
\put(99.5,25){\vector(1,0){1}}
\put(117,20){\line(0,1){10}}
\put(117,22.5){\vector(0,-1){1}}
\put(117,27.5){\vector(0,1){1}}
\put(112,25){\line(1,0){10}}
\put(114.5,25){\vector(1,0){1}}
\put(119.5,25){\vector(-1,0){1}}
\linethickness{0.5mm}
\put(7.25,10.25){\line(0,1){4.75}}
\put(7.5,10.25){\line(-1,0){5}}
\put(7.75,9.75){\line(0,-1){4.75}}
\put(7.5,9.75){\line(1,0){5}}
\thinlines
\put(27,5){\line(0,1){10}}
\put(22,10){\line(1,0){10}}
\linethickness{0.5mm}
\put(51.75,5){\line(0,1){10}}
\thinlines
\put(47,10){\line(1,0){10}}
\thinlines
\put(72,5){\line(0,1){10}}
\linethickness{0.5mm}
\put(67,9.75){\line(1,0){10}}
\linethickness{0.5mm}
\put(96.75,10){\line(0,1){5}}
\put(96.9,10){\line(-1,0){5}}
\thinlines
\put(97,9.75){\line(0,-1){5}}
\put(97,9.7){\line(1,0){4.7}}
\put(117,10.25){\line(0,1){4.75}}
\put(117,10.25){\line(-1,0){5}}
\linethickness{0.5mm}
\put(117.25,10){\line(0,-1){4.75}}
\put(117,10){\line(1,0){5}}
\put(5,-5){$w_1$}
\put(25,-5){$w_2$}
\put(50,-5){$w_3$}
\put(70,-5){$w_4$}
\put(95,-5){$w_5$}
\put(115,-5){$w_6$}
\end{picture}
\end{center}
\caption{The six allowed types of vertices
in terms of arrows (first row),
in terms of lines (second row), and their Boltzmann weights
(third row).}
\label{vertices}
\end{figure}
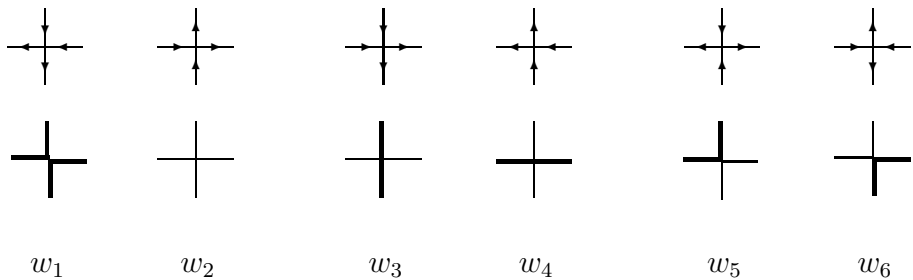

Let us start with recalling the formulation of the model.
The six-vertex model, which was originally proposed as
a model of two-dimensional ice,  is formulated on a square lattice
with arrows lying on edges, and
obeying the so called `ice-rule', namely, the
only admitted configurations are such that there are always two
arrows pointing away from, and two arrows pointing into, each lattice
vertex.
An equivalent and  graphically simpler description
of the configurations of the model can be given
in terms of lines flowing through the vertices: for each arrow
pointing downward or to the left, draw a thick line on the
corresponding
link. The six possible vertex states
and the Boltzmann weights $w_i$
assigned to each vertex according
to its state $i$ ($i=1,\dots,6$) are shown in Fig.~\ref{vertices}.
As it has been already stressed in the Introduction
only the homogeneous
version of the model, where the Boltzmann weights are site independent,
is considered here.

The DWBC are imposed on the $N\times N$ square lattice by fixing
the  direction of  all arrows on the boundaries in a specific way.
Namely, the vertical
arrows on the top and bottom
of the lattice point inward, while the horizontal arrows on the left
and right sides point outward. Equivalently, a generic configuration
of the model with  DWBC  can
be depicted by $N$ lines flowing from the upper boundary to the left one.
This line picture (besides taking into account the `ice rule'
in an automated way) is intuitively closer to ASMs
recalled in the next Section. A possible state of the model both in terms
of arrows and of lines is shown in Fig.~2.
\begin{figure}[t]
\unitlength=1mm
\begin{center}
\begin{picture}(30,30)
\put(0,5){\line(1,0){30}}
\put(5,5){\vector(-1,0){3.5}}
\put(10,5){\vector(-1,0){3.5}}
\put(15,5){\vector(-1,0){3.5}}
\put(20,5){\vector(-1,0){3.5}}
\put(25,5){\vector(-1,0){3.5}}
\put(25,5){\vector(1,0){3.5}}
\put(0,10){\line(1,0){30}}
\put(5,10){\vector(-1,0){3.5}}
\put(10,10){\vector(-1,0){3.5}}
\put(10,10){\vector(1,0){3.5}}
\put(15,10){\vector(1,0){3.5}}
\put(20,10){\vector(1,0){3.5}}
\put(25,10){\vector(1,0){3.5}}
\put(0,15){\line(1,0){30}}
\put(5,15){\vector(-1,0){3.5}}
\put(10,15){\vector(-1,0){3.5}}
\put(15,15){\vector(-1,0){3.5}}
\put(20,15){\vector(-1,0){3.5}}
\put(20,15){\vector(1,0){3.5}}
\put(25,15){\vector(1,0){3.5}}
\put(0,20){\line(1,0){30}}
\put(5,20){\vector(-1,0){3.5}}
\put(5,20){\vector(1,0){3.5}}
\put(15,20){\vector(-1,0){3.5}}
\put(15,20){\vector(1,0){3.5}}
\put(20,20){\vector(1,0){3.5}}
\put(25,20){\vector(1,0){3.5}}
\put(0,25){\line(1,0){30}}
\put(5,25){\vector(-1,0){3.5}}
\put(10,25){\vector(-1,0){3.5}}
\put(10,25){\vector(1,0){3.5}}
\put(15,25){\vector(1,0){3.5}}
\put(20,25){\vector(1,0){3.5}}
\put(25,25){\vector(1,0){3.5}}
\put(5,0){\line(0,1){30}}
\put(5.05,0){\vector(0,1){3.5}}
\put(5.05,5){\vector(0,1){3.5}}
\put(5.05,10){\vector(0,1){3.5}}
\put(5.05,15){\vector(0,1){3.5}}
\put(5.05,25){\vector(0,-1){3.5}}
\put(5.05,30){\vector(0,-1){3.5}}
\put(10,0){\line(0,1){30}}
\put(10.05,0){\vector(0,1){3.5}}
\put(10.05,5){\vector(0,1){3.5}}
\put(10.05,15){\vector(0,-1){3.5}}
\put(10.05,20){\vector(0,-1){3.5}}
\put(10.05,20){\vector(0,1){3.5}}
\put(10.05,30){\vector(0,-1){3.5}}
\put(15,0){\line(0,1){30}}
\put(15.05,0){\vector(0,1){3.5}}
\put(15.05,5){\vector(0,1){3.5}}
\put(15.05,10){\vector(0,1){3.5}}
\put(15.05,15){\vector(0,1){3.5}}
\put(15.05,25){\vector(0,-1){3.5}}
\put(15.05,30){\vector(0,-1){3.5}}
\put(20,0){\line(0,1){30}}
\put(20.05,0){\vector(0,1){3.5}}
\put(20.05,5){\vector(0,1){3.5}}
\put(20.05,10){\vector(0,1){3.5}}
\put(20.05,20){\vector(0,-1){3.5}}
\put(20.05,25){\vector(0,-1){3.5}}
\put(20.05,30){\vector(0,-1){3.5}}
\put(25,0){\line(0,1){30}}
\put(25.05,0){\vector(0,1){3.5}}
\put(25.05,10){\vector(0,-1){3.5}}
\put(25.05,15){\vector(0,-1){3.5}}
\put(25.05,20){\vector(0,-1){3.5}}
\put(25.05,25){\vector(0,-1){3.5}}
\put(25.05,30){\vector(0,-1){3.5}}
\put(12.5,-5){(a)}
\end{picture}
\qquad
\begin{picture}(30,30)
\put(0,5){\line(1,0){30}}
\put(0,10){\line(1,0){30}}
\put(0,15){\line(1,0){30}}
\put(0,20){\line(1,0){30}}
\put(0,25){\line(1,0){30}}
\put(5,0){\line(0,1){30}}
\put(10,0){\line(0,1){30}}
\put(15,0){\line(0,1){30}}
\put(20,0){\line(0,1){30}}
\put(25,0){\line(0,1){30}}
\linethickness{0.5mm}
\put(0,5){\line(1,0){25.25}}
\put(25,30){\line(0,-1){25.25}}
\put(0,10){\line(1,0){10.25}}
\put(20,30){\line(0,-1){15.25}}
\put(20.25,14.75){\line(-1,0){10}}
\put(10.25,9.75){\line(0,1){5.25}}
\put(0,15.25){\line(1,0){10}}
\put(15,30){\line(0,-1){10.25}}
\put(15,20){\line(-1,0){5.5}}
\put(9.75,20){\line(0,-1){5}}
\put(0,20){\line(1,0){5.25}}
\put(10,30){\line(0,-1){5.25}}
\put(10.25,24.75){\line(-1,0){5}}
\put(5.25,25){\line(0,-1){5.25}}
\put(0,25.25){\line(1,0){5}}
\put(4.75,30){\line(0,-1){5}}
\put(12.5,-5){(b)}
\end{picture}
\end{center}
\caption{One of the possible configurations of the model with DWBC, in
the case$N=5$:
(a) in terms of arrows; (b) in terms of lines.}
\label{dwbcgrid}
\end{figure}
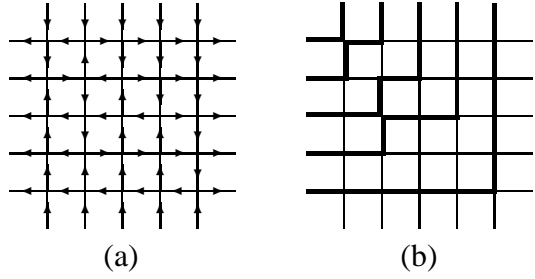

The partition function is defined, as usual, as a
sum over all possible arrow configurations, compatible with
the imposed DWBC, each configuration being assigned its Boltzmann weight,
given as the product of all the corresponding vertex weights,
\begin{equation}
Z_N=\sum_{\substack{\text{arrow configurations}\\
\text{with DWBC}}}^{}\
\prod_{i=1}^{6}w_i^{n_i}\;.
\end{equation}
Here $n_i$ denotes the number of vertices in the state $i$
in each arrow configuration ($n_1+\dots+n_6=N^2$).

The six-vertex model with DWBC can be considered,
with no loss of generality,
with its weights invariant under the simultaneous reversal
of all arrows,
\begin{equation}
w_1=w_2=:a,\qquad
w_3=w_4=:b,\qquad
w_5=w_6=:c.
\end{equation}
Under different choices of Boltzmann weights
the six-vertex model exhibits different  behaviors,
according to the value of the parameter $\Delta$ defined as
\begin{equation}
\Delta=\frac{a^2+b^2-c^2}{2ab}.
\end{equation}
It is well known that
there are three physical regions or phases for the six-vertex model:
the ferroelectric phase, $\Delta>1$;
the anti-ferroelectric phase, $\Delta<-1$;
and, the disordered phase,
$-1<\Delta<1$. In the present paper we shall discuss only
some particular cases, with values of the Boltzmann weights
that correspond to the disordered phase. A convenient parametrization
of the Boltzmann weights in this phase is
\begin{equation} \label{sin}
a=\sin(\lambda+\eta),\qquad
b=\sin(\lambda-\eta),\qquad
c=\sin 2\eta.
\end{equation}
With this choice one has $\Delta=\cos 2\eta$. The parameter $\lambda$
is the so-called spectral parameter
and $\eta$ is the crossing parameter.
The physical requirement of positive Boltzmann weights, in the disordered
regime, restricts the values of the  crossing and  spectral parameters
to $0<\eta<\pi/2$ and $\eta<\lambda<\pi/2-\eta$.

An exact representation for the partition function was
obtained in Ref.~\cite{ICK-92}.
When the weights are parameterized according to \eqref{sin}
such representation reads
\begin{align}\label{Z=detZ}
Z_N=\frac{[\sin(\lambda-\eta)\sin(\lambda+\eta)]^{N^2}}
{\prod\limits_{k=1}^{N-1}(k!)^2}\;
{\det}\,\mathcal{Z}
\end{align}
where $\mathcal{Z}$ is an $N\times N$ matrix with entries
\begin{equation}\label{Zmat}
\mathcal{Z}_{jk}=
\frac{\partial^{j+k}}{\partial\lambda^{j+k}}\,
\frac{\sin2\eta}{\sin(\lambda-\eta)\sin(\lambda+\eta)}\;.
\end{equation}
Here and in the following we use the convention that indices
of $N\times N$ matrices run
over the values $j,k=0,1,\dots,N-1$.

This formula for the partition function has been obtained
as the homogeneous limit of a  more
general formula for a partially inhomogeneous
six-vertex model with DWBC. The inhomogeneous model,
with site-dependent weights, is defined by introducing
two sets of spectral parameters $\{\lambda_j\}_{j=1}^N$ and
$\{\nu_k\}_{k=1}^N$, such that the weights of the vertex lying at the
intersection of the $j$-th column with the $k$-th row depend on
$\lambda_j-\nu_k$ rather than simply on  $\lambda$, still
through formulae \eqref{sin}. The inhomogeneous model,
though apparently more
complicate, can be fruitfully investigated through  the
Quantum Inverse Scattering Method,  see papers \cite{K-82,I-87,ICK-92}
and book \cite{KBI-93} for details. As a  result, the partition
function of the inhomogeneous model is represented in terms of
certain determinant
formula which, however, requires  great caution in the study of its
homogeneous limit, $\nu_k\to 0$ and $\lambda_j\to\lambda$, since
in this limit the determinant possess $N^2-N$ zeros
that are cancelled by the same number of singularities coming
from the pre-factor. A recipe for taking such a
limit was explained in detail in Ref.~\cite{ICK-92}
where formula \eqref{Z=detZ} was originally obtained.
Subsequently, formula \eqref{Z=detZ} was used in papers
\cite{KZj-00,Zj-00} to investigate the thermodynamic limit, $N\to\infty$,
of the partition function. In these  studies the
Hankel nature of the determinant appearing in \eqref{Z=detZ}, a
natural outcome of the  homogeneous limit procedure, was exploited
through its relation with   the  Toda chain differential equation
or with the random matrix partition function.

The aim of this paper is to explain how this Hankel determinant formula
(and its analogue for the boundary correlator, given below)
can also be used  in application to some well-known problems
of combinatorics,
such as enumerations (and the so-called `refined' enumerations)
of ASMs.
These problems and the known
results are reviewed in the next Section.
It is to be mentioned
that though in these combinatorial  problems one deals
in fact with the homogeneous
six-vertex model with DWBC, the Hankel determinant formula for the
partition function was never  discussed previously in this context.
Instead, the more complicated determinant formula
for the inhomogeneous model partition function was used
in the combinatorial proofs. In these  proofs,  the homogeneous limit
can be regarded, in fact,  as the most complicated part on
the way to the result.
Consequently, one can expect that extracting relevant information
directly from the homogeneous model should
be technically much simpler. We will show that this is indeed the case
since some standard  classical orthogonal polynomials can be naturally
related (for some  particular  choices of the parameters) to the Hankel
determinant in \eqref{Z=detZ}.

In addition to the partition function, we shall discuss here
also boundary one point correlation functions.
In general,  two kinds of one point correlation functions
can be considered in the six-vertex model: the first one (`polarization')
is the probability to find an arrow on a given edge in a particular
state,
while the second one is the probability to find a given vertex in
some state $i$.
If one restricts to edges or vertices adjacent to the boundary,
then such correlators are called boundary correlators.
Following the notations of paper \cite{BPZ-02}, where these boundary
correlators were studied, let $G_N^{(r)}$ denote the probability
that an arrow on the last column and between the $r$-th and the $r+1$-th
rows (enumerated from the bottom) points upward
(or, in the line language,
that there is no thick line on this edge), and let $H_N^{(r)}$ denote
the probability that the $r$-th vertex of the last column is in the
state $i=5$ (or that the thick line flows from the top to the left),
see Figs.~\ref{vertices} and \ref{dwbcgrid}.
The first correlator, $G_N^{(r)}$, is, in fact, the
boundary polarization, whose interpretation is more direct from
a physical point of view, while
the second one,  $H_N^{(r)}$, is closely related to the refined
enumerations of ASMs. It is easy to see that,
due to DWBC, the two correlators are related to each other as follows
\begin{equation}\label{GviaH}
G_N^{(r)}= H_N^{(r)} + H_N^{(r-1)} +\cdots + H_N^{(1)}.
\end{equation}

In Ref.~\cite{BPZ-02} both correlators were computed using Quantum
Inverse Scattering Method for the inhomogeneous six-vertex model.
In the homogeneous limit, which is the situation we are interested
in here, determinant formulae generalizing \eqref{Z=detZ} were
found for these correlators. For instance, for $H_N^{(r)}$,
the following expression was derived
\begin{equation} \label{HNM}
H_N^{(r)}=\frac{(N-1)!\,\sin 2\eta}{
\bigl[\sin(\lambda-\eta)\bigr]^r\bigl[\sin(\lambda+\eta)\bigr]^{N-r+1}}
\frac{\det\,\mathcal{H}}{\det\, \mathcal{Z}}
\end{equation}
where the $N\times N$ matrix $\mathcal{H}$ differs
from $\mathcal{Z}$ only in the last column,
\begin{equation} \label{Hmat}
\mathcal{H}_{jk} =
\begin{cases}
\mathcal{Z}_{jk} &\text{for}\quad k=0,\ldots, N-2
\\
\dfrac{\partial^j}{\partial\eps^j}\,
\dfrac{\bigl[\sin\eps\bigr]^{r-1}\bigl[\sin(\eps-2\eta)\bigr]^{N-r}}{
\bigl[\sin(\eps+\lambda-\eta)\bigr]^{N-1}}
\Bigg|_{\eps=0} &\text{for}\quad k=N-1
\end{cases}.
\end{equation}
A similar expression is valid for $G_N^{(r)}$ as well.
In what follows we shall focus on $H_N^{(r)}$; the results for
$G_N^{(r)}$ will follow immediately from relation \eqref{GviaH}.
From the DWBC it immediately follows that $G_N^{(N)}=1$, and
therefore, from \eqref{GviaH},  correlator $H_N^{(r)}$
has to satisfy
\begin{equation}\label{normcond}
\sum_{r=1}^N \ H_N^{(r)} = 1 \,.
\end{equation}
In what follows this normalization condition
will be used in application to the generating function of
$H_N^{(r)}$.

\section{Alternating Sign Matrices}

An alternating sign matrix (ASM)  is a matrix of $1$'s, $0$'s and $-1$'s
such that in each row and in each column \emph{(i)} all nonzero entries
alternate  in sign, and  \emph{(ii)}  the
first and the last nonzero entries  are  1. An example of
such matrix is
\begin{equation}\label{ASM}
\begin{pmatrix}
0 & 1 & 0 & 0 & 0 \\
1 & -1 & 1 & 0 & 0 \\
0 & 0 & 0 & 1 & 0 \\
0 & 1 & 0 & 0 & 0 \\
0 & 0 & 0 & 0 & 1
\end{pmatrix} .
\end{equation}
There are many nice results concerning ASMs,
for a review, see book \cite{Br-99}.
Many of these results have been first formulated as conjectures
which were subsequently proved by different methods.

The most celebrated result concerns
the total number $A(N)$ of $N\times N$ ASMs.
It was conjectured in papers \cite{MRR-82,MRR-83}
and proved in papers \cite{Z-96a,Ku-96} that
\begin{equation}\label{An}
A(N)=\prod_{k=1}^{N}\frac{(3k-2)!\, (k-1)!}{(2k-1)!\,
(2k-2)!} =
\prod_{k=1}^{N}   \frac{(3k-2)!}{(2N-k)!}.
\end{equation}
Other results concern the weighted enumerations or the so-called
$x$-enumerations of ASMs.
In $x$-enumeration the matrices are counted
with a weight $x^k$ where $k$ is the total number of
$-1$ entries in a matrix (the number $x$ here should not be
confused with the
variable $x$ widely used in the following  Sections).
The number of $x$-enumerated ASMs is denoted traditionally as
$A(N;x)$. The extension of the $x=1$ result above
to the case of generic $x$ is not known, but
for a few nontrivial cases, namely $x=2$ and $x=3$, closed expressions
for $x$-enumerations are known (the case $x=0$ is trivial
since assigning a vanishing weight to each
$-1$ entry restricts  the enumeration to the sole permutation matrices:
$A(n;0)=n!$). The result in the case $x=2$,
related to the domino tilings of Aztec diamond
\cite{EKLP-92a,EKLP-92b,JPS-98}, have been obtained in \cite{MRR-83},
and reads
\begin{equation} \label{An2}
A(N;2)=2^{N(N-1)/2}.
\end{equation}
The answer in the case of $3$-enumeration,
again conjectured in \cite{MRR-82,MRR-83}, was subsequently \cite{Ku-96}
proved to be
\begin{equation}\label{An3}
A(2m+1;3)= 3^{m(m+1)}
\prod_{k=1}^{m}\! \left[\frac{(3k-1)!}{(m+k)!}\right]^2,
\qquad
A(2m+2;3)= 3^m \frac{(3m+2)!\, m!}{\bigl[(2m+1)!\bigr]^2}\; A(2m+1;3).
\end{equation}

Another  class  of results concerns the so-called
refined enumerations of ASMs. In the refined enumeration
one counts the number of $N\times N$ ASMs
with their sole $1$ of the last column
at the $r$-th entry. The refined enumeration can
be naturally extended to be also an $x$-enumeration.
The standard notation for the refined $x$-enumeration
is $A(N,r;x)$; in the case $x=1$ one writes simply $A(N,r)$
just like $A(N)$ for the total number of ASMs. It was again
conjectured in \cite{MRR-82,MRR-83}, and proved in \cite{Z-96b} that
refined enumeration of ASMs is given by
\begin{equation}\label{Anr}
A(N,r)
=\frac{\binom{N+r-2}{N-1}\binom{2N-1-r}{N-1}}{\binom{3N-2}{N-1}}\;
A(N).
\end{equation}
In the case of the refined $2$-enumeration it has been shown
in \cite{MRR-83,EKLP-92a,EKLP-92b,JPS-98} that
\begin{equation}\label{Anr2}
A(N,r;2)=\frac{1}{2^{N-1}}\binom{N-1}{r-1}\, A(N;2).
\end{equation}
The case of the refined $3$-enumeration appears
to be much more complicate.
Direct computer enumeration does not suggest any factorized form
and no conjecture concerning  $A(N,r;3)$ has ever been proposed.
Recently, Stroganov obtained a certain representation
for the corresponding generating function \cite{S-03},
allowing in principle a recursive computation of the numbers $A(N,r;3)$.

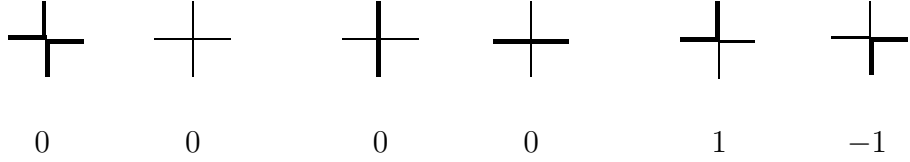
\begin{figure}[t]
\unitlength=1mm
\begin{center}
\begin{picture}(124,10)
\linethickness{0.5mm}
\put(7.25,10.25){\line(0,1){4.75}}
\put(7.5,10.25){\line(-1,0){5}}
\put(7.75,9.75){\line(0,-1){4.75}}
\put(7.5,9.75){\line(1,0){5}}
\thinlines
\put(27,5){\line(0,1){10}}
\put(22,10){\line(1,0){10}}
\linethickness{0.5mm}
\put(51.75,5){\line(0,1){10}}
\thinlines
\put(47,10){\line(1,0){10}}
\thinlines
\put(72,5){\line(0,1){10}}
\linethickness{0.5mm}
\put(67,9.75){\line(1,0){10}}
\linethickness{0.5mm}
\put(96.75,10){\line(0,1){5}}
\put(96.9,10){\line(-1,0){5}}
\thinlines
\put(97,9.75){\line(0,-1){5}}
\put(97,9.7){\line(1,0){4.7}}
\put(117,10.25){\line(0,1){4.75}}
\put(117,10.25){\line(-1,0){5}}
\linethickness{0.5mm}
\put(117.25,10){\line(0,-1){4.75}}
\put(117,10){\line(1,0){5}}
\put(6,-5){$0$}
\put(26,-5){$0$}
\put(51,-5){$0$}
\put(71,-5){$0$}
\put(96,-5){$1$}
\put(114,-5){$-1$}
\end{picture}
\end{center}
\caption{Vertex states---ASM's entries correspondence.}
\label{aentries}
\end{figure}

The most direct way to recover all the previous results is based
on a nice bijective correspondence between ASMs and
six-vertex model, which has been pointed out in
\cite{EKLP-92a,EKLP-92b,RR-86},
and applied for the first time in \cite{Ku-96}, namely,
that to each single $N\times N$ ASM
corresponds one and only one arrow configuration of  the
six-vertex model on the $N\times N$ square lattice with DWBC.
The correspondence between matrix entries and vertices is depicted in
Fig.~\ref{aentries}.
As an example, matrix \eqref{ASM}
corresponds to the configuration of Fig.~\ref{dwbcgrid} and
vice versa.

As an immediate consequence of this correspondence, ASM enumeration
is exactly given by the
partition function of square ice, when all
vertex weights are set equal to unity. More generally,
the number of $-1$'s in a given ASM
being equal to the number of vertices of type 6 (see Fig.~\ref{aentries}),
and the number of vertex of type 5 and 6 being  constrained by
the condition $n_5-n_6 =N$, we readily get
\begin{equation}\label{AnZn}
A(N;x)
=(1-x/4)^{-N^2/2}\, x^{-N/2}\,
Z_N\Big|_{
\begin{subarray}{l}
\lambda=\pi/2\\
\eta=\arcsin(\sqrt{x}/2)
\end{subarray}
}.
\end{equation}
Therefore, $x$-enumeration of ASM corresponds to the computation of
the partition function of square ice on the subset of parameters space
given by $a=b$. In this correspondence,
values of $x$ belonging to the interval $(0,4)$  corresponds to the
disordered regime of the model, $-1<\Delta<1$.

This nice correspondence can be further extended to
the refined $x$-enumeration of ASMs.
In the language of square ice, the ratio  $A(N,r;x)/A(N;x)$ can
be rephrased as the probability of finding the unique vertex of type 5
on the right  boundary at the
$r$-th site, which is exactly the definition  of the
boundary correlator $H_N^{(r)}$. Explicitly, one has
\begin{equation}\label{AnrHnr}
\frac{A(N,r;x)}{A(N;x)}
=H_N^{(r)}\Big|_{
\begin{subarray}{l}
\lambda=\pi/2\\
\eta=\arcsin(\sqrt{x}/2)
\end{subarray}
}.
\end{equation}
Thus, being able to compute the partition function and
the boundary correlator for some particular choice of parameters,
one immediately obtains  $x$-enumerations and refined $x$-enumerations
of ASMs respectively, for some corresponding values of x.

\section{The partition function and enumerations of ASMs}\label{xenume}

\subsection{Preliminaries}

From representation \eqref{Z=detZ}, it is evident that
the evaluation of the  the partition function in the homogeneous case
essentially reduces, modulo a trivial pre-factor, to the
calculation of the determinant of the $N\times N$ matrix $\mathcal{Z}$,
which  is a Hankel matrix, whose entries do not depend on $N$.
There is  a standard  method to treat such determinants, which
is based on the theory of orthogonal polynomials \cite{S-75,E-81},
and has proven to be quite powerful when some assumption are verified.

Let us assume that the entries  of our Hankel matrix are written
in the canonical  form
\begin{equation}\label{moments}
\mathcal{Z}_{jk} = \int_{-\infty}^{\infty} x^{j+k}
\mu(x)\, \mathrm{d}x.
\end{equation}
Let us moreover suppose that there exist a complete set of polynomials
$\{p_n(x)\}_{n=0}^\infty$ subject to the orthogonality condition
\begin{equation}
\int_{-\infty}^{\infty}
p_j(x)\, p_k(x)\,\mu(x)\,\mathrm{d}x
= h_j \delta_{jk}\;.
\end{equation}
Then, denoting by $\kappa_n$ the leading coefficient of $p_n(x)$,
\begin{equation}
p_n(x)=\kappa_n x^n+\dots\;,\qquad \kappa_n \ne 0\;,
\end{equation}
and using standard properties of  determinants and of
orthogonal polynomials, one obtains for
the determinant of the $N\times N$ matrix $\mathcal{Z}$
the following formula
\begin{align}
{\det} \mathcal{Z} &= \det \left[
\int \frac{1}{\kappa_j \kappa_k}
p_j(x) p_k(x) \mu(x) \mathrm{d}x
\right]_{j,k=0}^{N-1} \\
&=\prod_{n=0}^{N-1} \frac{h_n}{\kappa_n^2}\;.
\label{detZ}\end{align}
Obviously, this formula may turn out useful provided that
the set of polynomials which are orthogonal with respect to
the weight $\mu(x)$ can be identified.

In the case of matrix $\mathcal{Z}$ with entries \eqref{Zmat}
one can easily obtain the weight $\mu(x)$ using the representation
\begin{equation}\label{mu}
\frac{\sin2\eta}{\sin(\lambda-\eta)\sin(\lambda+\eta)}
=\int_{-\infty}^{\infty}
\mathrm{e}^{(\lambda -\pi /2) x}\,
\frac{\sinh\eta x}{\sinh \frac{\pi}{2}x}
\,\mathrm{d}x
\end{equation}
which is valid for $\lambda$ and $\eta$ corresponding to the disordered
regime \cite{Zj-00}.
Unfortunately,
appropriate polynomials are not available in general and the
previous scheme cannot be fulfilled
for generic values of $\lambda$ and $\eta$.
However, for some very particular values of these parameters
the appropriate orthogonal polynomials appear to be known, and have just
to be suitably chosen in the framework of Askey scheme of
hypergeometric orthogonal polynomials \cite{KS-98}.
\begin{figure}[t]
\begin{center}
\begin{pspicture}(-1,0)(5,5)
\psset{arrowscale=1.5}
\psline{->}(0,0)(4.8,0)
\psline{->}(0,0)(0,4.8)
\psline(0,3)(1.5,4.5)
\psline(3,0)(4.5,1.5)
\psline(0,3)(3,0)
\psline[linestyle=dashed,linewidth=.5pt](0,0)(4,4)
\rput(-.2,-.4){$0$}
\rput(3,-.4){$1$}
\rput(-.4,3){$1$}
\rput(4.5,-.4){$a/c$}
\rput(-.5,4.5){$b/c$}
\psarc[linewidth=1.8pt](0,0){3}{0}{90}
\psdot[dotscale=1.5](2.12,2.12)
\psline[linestyle=dotted](0,2.12)(2.12,2.12)(2.12,0)
\psdot[dotscale=1.5](3,3)
\psline[linestyle=dotted](0,3)(3,3)(3,0)
\psdot[dotscale=1.5](1.75,1.75)
\psline[linestyle=dotted](0,1.75)(1.75,1.75)(1.75,0)
\rput(1.7,-.4){$\frac{1}{\sqrt{3}}$}
\rput(2.2,-.4){$\frac{1}{\sqrt{2}}$}
\rput(2.5,3.8){$\mathsf{D}$}
\rput(.5,4.2){$\mathsf{F}$}
\rput(4.2,.5){$\mathsf{F}$}
\rput(.6,1.2){$\mathsf{AF}$}
\end{pspicture}
\end{center}
\caption{The phase diagram of the model,
with ferroelectric ($\mathsf{F}$),
antiferroelectric ($\mathsf{AF}$) and disordered ($\mathsf{D}$) phases,
separated by the solid lines.
The three considered cases, all belonging to the
disordered phase, are shown in bold: the free fermion line,
and the three points corresponding to $1$-, $2$- and $3$-enumeration
of ASMs.}
\label{diagramm}
\end{figure}
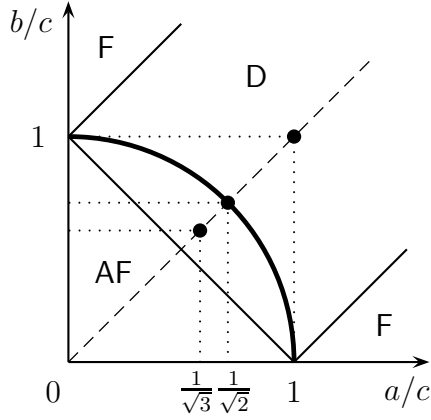
To be precise, there are essentially three cases, indicated
in the phase diagram of the model, see Fig.~\ref{diagramm},
which fit into the scheme:
{\it i)} the so-called `free fermion' line $\eta=\pi/4$ and
$\pi/4<\lambda<3\pi/4$; {\it ii)}
the $\Delta=1/2$ symmetric point (or `ice point')
$\eta=\pi/6$ and $\lambda=\pi/2$; and {\it iii)}  the $\Delta=-1/2$
symmetric point $\eta=\pi/3$ and $\lambda=\pi/2$.
With `symmetric' here  we mean that these points lie on
the line $a=b$. The polynomials corresponding to these three cases are
Meixner-Pollaczek polynomials, Continuous Hahn polynomials and
Continuous Dual Hahn polynomials, respectively.
In the rest of this Section we give details of computation
for  each of these cases.

It is worth mentioning that all previously listed choices of parameters,
for which the Hankel determinant appearing  in \eqref{Z=detZ}
turns out to be related to  some set of  classical orthogonal
polynomials,
exactly cover $1$-, $2$-  and  $3$-enumerations of ASMs.
On the other hand, the fact that no set of polynomials in the
Askey scheme corresponds to some other choice of the parameter
$\eta$ (even with  $\lambda$ set equal to $\pi/2$) is likely
to be deeply
related to the lack of factorizable formulae for
$x$-enumerations of ASMs other than  for $x=1,2,3$.

\subsection{The free fermion line}

We shall start with the $\eta=\pi/4$ case which is technically
the simplest one,  though the parameter $\lambda$ is not fixed,
$\pi/4<\lambda<3\pi/4$. In this case the orthogonality weight
$\mu(x)$ can be written as
\begin{equation}
\mu(x)=\mathrm{e}^{(\lambda-\pi/2) x}\,
\frac{\sinh \frac{\pi}{4} x}{\sinh \frac{\pi}{2}x}
= \frac{\mathrm{e}^{(\lambda-\pi/2) x}}{2\cosh\frac{\pi}{4}x}=
\frac{\mathrm{e}^{(\lambda-\pi/2) x}}{2\pi}
\bigg|\Gamma\biggl(\frac{1}{2}+\mathrm{i}\frac{x}{4}\biggr)\bigg|^2.
\end{equation}
Comparing this formula with the orthogonality condition
for Meixner-Pollaczek polynomials \cite{KS-98}
\begin{equation}\label{MPoc}
\frac{1}{2\pi}\int_{-\infty}^{\infty}
\MP{m}{\alpha}{x}{\phi} \MP{n}{\alpha}{x}{\phi}
|\Gamma(\alpha+\mathrm{i} x)|^2\,
\mathrm{e}^{(2\phi-\pi)x}\,\mathrm{d} x
=\frac{\Gamma(n+2\alpha)}{(2\sin\phi)^{2\alpha}\,n!}\,\delta_{nm}
\end{equation}
where
\begin{equation}\label{MPdef}
\MP{n}{\alpha}{x}{\phi}
=\frac{(2\alpha)_n}{n!}\,\mathrm{e}^{\mathrm{i} n\phi}
\F{-n}{\alpha+\mathrm{i}x}{2\alpha}{1-\mathrm{e}^{-2\mathrm{i}\phi}}
\end{equation}
we find that the polynomials related to the Hankel determinant
in this case are
\begin{equation}\label{poly2}
p_n(x)=\MP{n}{1/2}{\frac{x}{4}}{2\lambda-\frac{\pi}{2}}.
\end{equation}
Using, for convenience, the parameter $\phi=2\lambda-\pi/2\in(0,\pi)$
instead of $\lambda$, we have
\begin{equation}
h_n=\frac{2}{\sin\phi},\qquad
\kappa_n=\frac{(\sin\phi)^n}{2^n\, n!}.
\end{equation}
Inserting these expressions into \eqref{detZ},
we straightforwardly obtain the following result
for the partition function
\begin{equation}
Z_N\Big|_{\eta=\pi/4} =
\frac{(\sin\phi)^{N^2}}{2^{N^2}\prod_{k=0}^{N-1} (k!)^2}
\prod_{n=0}^{N-1}
\frac{2}{\sin\phi} \left[\frac{2^n n!}{(\sin\phi)^n}\right]^2=1.
\end{equation}
Using Eqn.~\eqref{AnZn}, we immediately recover
formula \eqref{An2} for $2$-enumeration of ASMs.

\subsection{The ice point and the total number of ASMs}

At the ice point, or $\Delta=1/2$ symmetric point,
the computation is very similar. For $\eta=\pi/6$ and $\lambda=\pi/2$
the orthogonality weight reads
\begin{equation}\label{mu-one}
\mu(x)=\frac{\sinh\frac{\pi}{6}x}{\sinh\frac{\pi}{2} x}
=\frac{1}{4 \pi^2}
\bigg|\Gamma\biggl(\frac{1}{3}+i\frac{x}{6}\biggr)
\Gamma\biggl(\frac{2}{3}+i\frac{x}{6}\biggr)\bigg|^2.
\end{equation}
where the triplication formula for the $\Gamma$-function
\begin{equation}\label{triple}
\Gamma(3x)=\frac{3^{3x-1/2}}{2\pi}\; \Gamma(x)\,
\Gamma\biggl(x+\frac{1}{3}\biggr)\,
\Gamma\biggl(x+\frac{2}{3}\biggr)
\end{equation}
has been used.
The orthogonality condition for Continuous Hahn polynomials is
\begin{multline}\label{ortCHP}
\frac{1}{2\pi} \int_{-\infty}^{\infty}
p_n(x;a,b,c,d) p_m(x;a,b,c,d)\,
\Gamma(a+\mathrm{i} x)\,\Gamma(b+\mathrm{i} x)\,
\Gamma(c-\mathrm{i} x)\,\Gamma(d-\mathrm{i} x)
\,\mathrm{d}x
\\
=\frac{\Gamma(n+a+c)\,\Gamma(n+a+d)\,\Gamma(n+b+c)\,\Gamma(n+b+d)}
{(2n+a+b+c+d-1)\,\Gamma(n+a+b+c+d-1)\,n!}\, \delta_{nm}
\end{multline}
and the polynomials are given by
\begin{equation}
p_n(x;a,b,c,d)
=\mathrm{i}^n
\frac{(a+c)_n(a+d)_n}{n!}
\Fthreetwo{-n}{n+a+b+c+d-1}{a+\mathrm{i} x}{a+c}{a+d}{1}.
\end{equation}
Orthogonality condition \eqref{ortCHP} is valid if
the parameters $a,b,c,d$ satisfy $\Re(a,b,c,d)>0$, $a=\bar c$ and
$b=\bar d$. Comparing \eqref{mu-one} with \eqref{ortCHP} we are naturally
led to the choice of parameters $a=c=1/3$ and $b=d=2/3$. Hence,
the appropriate polynomials to be associated to the Hankel determinant
in this case are
\begin{equation}\label{poly1}
p_n(x)=
p_n\biggl(\frac{x}{6};\frac{1}{3},\frac{2}{3},
\frac{1}{3},\frac{2}{3}\biggr)
=\mathrm{i}^n (2/3)_n
\,\Fthreetwo{-n}{n+1}{1/3+\mathrm{i} x/6}{2/3}{1}{1}.
\end{equation}
For the normalization constant and the leading coefficient we have
the expressions
\begin{equation}
h_n=\frac{2 (3n+1)!}{(2n+1)\, 3^{3n+1/2}\, n!},\qquad
\kappa_n=\frac{(2n)!}{6^n\, (n!)^2}
\end{equation}
where the triplication formula, Eqn.~\eqref{triple}
has been used to simplify the normalization constant $h_n$.

Substituting the obtained values of $h_n$ and $\kappa_n$ in
expression \eqref{detZ}
for the determinant, taking into account the value of the prefactor
in \eqref{Z=detZ} for $\lambda=\pi/2$ and $\eta=\pi/6$,
and cancelling whatever possible, we arrive to the following value
for the ice point partition function
\begin{equation}
Z_N\Big|_{
\begin{subarray}{l}
\lambda=\pi/2\\
\eta=\pi/6
\end{subarray}
}
=\biggl(\frac{\sqrt{3}}{2}\biggr)^{N^2}\prod_{n=0}^{N-1}
\frac{(3n+1)!\, n!}{(2n)!\,(2n+1)!}.
\end{equation}
The product expression here gives exactly the total number of ASMs,
$A(N)$, since by formula \eqref{AnZn} the first factor
relates $A(N)$ with the partition function,
\begin{equation}
A(N)
=(3/4)^{-N^2/2} \,
Z_N\Big|_{
\begin{subarray}{l}
\lambda=\pi/2\\
\eta=\pi/6
\end{subarray}
}\;.
\end{equation}
Thus, we have easily recovered the celebrated result, Eqn.~\eqref{An},
for ASMs enumeration directly from the Hankel determinant formula
\eqref{Z=detZ}.

\subsection{The $\Delta=-1/2$ symmetric point and
$3$-enumeration of ASMs}\label{sec.3en}

In this case, $\lambda=\pi/2$ and $\eta=\pi/3$, the weight $\mu(x)$
can be rewritten, using \eqref{triple}, in the form
\begin{equation}\label{mu-three}
\mu(x)=\frac{\sinh\frac{\pi}{3} x}{\sinh\frac{\pi}{2} x}
=
\frac{1}{8 \pi^2}
\bigg|\frac{
\Gamma(\mathrm{i}\frac{x}{6})
\Gamma(\frac{1}{3}+\mathrm{i}\frac{x}{6})
\Gamma(\frac{2}{3}+\mathrm{i}\frac{x}{6})
}{\Gamma(\mathrm{i}\frac{x}{3})
}
\bigg|^2.
\end{equation}
This expression recalls the weight for
Continuous Dual Hahn polynomials $S_n(x^2;a,b,c)$, which are defined by
\begin{equation}
S_n(x^2;a,b,c)=(a+b)_n (a+c)_n
\Fthreetwo{-n}{a+\mathrm{i}x}{a-\mathrm{i}x}{a+b}{a+c}{1},
\end{equation}
and, for  real and
nonnegative values of parameters $a,b,c$, satisfy
the orthogonality condition
\begin{multline}\label{ortCDHP}
\frac{1}{2\pi}
\int_{0}^\infty
S_m(x^2;a,b,c) S_n(x^2;a,b,c)
\bigg|\frac{\Gamma(a+\mathrm{i}x)
\Gamma(b+\mathrm{i}x)\Gamma(c+\mathrm{i}x)
}{\Gamma(2\mathrm{i}x)}\bigg|^2
\mathrm{d}x
\\
=\Gamma(n+a+b)\Gamma(n+a+c)\Gamma(n+b+c)\, n!\,\delta_{nm}\;.
\end{multline}
At a  first glance, however,  Continuous Dual Hahn polynomials do
not seem appropriate for  the evaluation of the Hankel determinant:
on one hand, these  are polynomials in $x^2$ rather than in $x$, and
on the other hand the integration domain in \eqref{ortCDHP}
is restricted to the positive half-axis.
These inconvenience may  nevertheless easily be circumvented,
thanks to the following nice feature of
matrix $\mathcal{Z}$, holding for $\lambda=\pi/2$ and generic
$\eta$: whenever the sum of indices $j+k$ is odd, the corresponding entry
$\mathcal{Z}_{jk}$ vanishes. In other words,  the matrix
$\mathcal{Z}$ at $\lambda=\pi/2$ exhibit a chessboard pattern
of vanishing and non vanishing entries. As can be seen from the Laplace
formula for the determinant of the sum of two matrices, this chessboard
structure immediately implies that
the determinant of such  matrix always factorizes onto two determinants
of smaller matrices with no vanishing entries.

To be precise, let us denote
\begin{equation}\label{Deltan}
D_N:=\det\mathcal{Z}\bigg|_{\substack{\lambda=\pi/2\\ \eta=\pi/3}}
=\det\left[\int_{-\infty}^{\infty}
x^{j+k}\,\mu(x)\,\mathrm{d}x \right]_{j,k=0}^{N-1}
\end{equation}
where $\mu(x)$ is given by \eqref{mu-three}. Since the orthogonality
weight is even, $\mu(x)=\mu(-x)$,
one has
\begin{equation}
\int_{-\infty}^{\infty}
x^{j+k}\,\mu(x)\,\mathrm{d}x
=
\begin{cases}
2\displaystyle \int_{0}^{\infty}
x^{j+k}\,\mu(x)\,\mathrm{d}x & \text{if}\,j+k\,\text{is even}\\
0& \text{if}\,j+k\,\text{is odd}
\end{cases}
\end{equation}
and consequently the following factorization arises
\begin{equation}\label{D=DD}
D_{2m}^{}=D^{(0)}_{m} D^{(1)}_{m},\qquad
D_{2m+1}^{}=D^{(0)}_{m+1} D^{(1)}_{m}.
\end{equation}
Here
$D_m^{(0)}$ and $D_m^{(1)}$ are determinants of
$m\times m$ Hankel matrices, built only
from the even moments of the weight $\mu(x)$,
\begin{equation}
D^{(\sigma)}_m=
\det\left[\int_0^{\infty}x^{2(j+k)}\,\mu^{(\sigma)}(x)
\,\mathrm{d}x\right]_{j,k=0}^{m-1},\qquad
\sigma=0,1
\end{equation}
where
\begin{equation}\label{mu0-mu1}
\mu^{(0)}(x)=2 \mu(x),\qquad
\mu^{(1)}(x)=2 x^2 \mu(x).
\end{equation}
Let $p_n^{(\sigma)}(x)$ be the polynomials subject to the orthogonality
condition
\begin{equation}
\int_{0}^\infty p_j^{(\sigma)}(x^2) p_k^{(\sigma)}(x^2)
\mu^{(\sigma)}(x)\, \mathrm{d}x = h_j^{(\sigma)} \delta_{jk},
\end{equation}
then, in analogy
with Eqn.~\eqref{detZ}, we have the following formula
\begin{align}\label{detD}
D_m^{(\sigma)}&=
\det\left[\int_0^{\infty}
\frac{1}{\kappa_j^{(\sigma)}\kappa_{k}^{(\sigma)}}
\,p_j^{(\sigma)}(x^2) \, p_k^{(\sigma)}(x^2)
\,\mu^{(\sigma)}(x)
\,\mathrm{d}x\right]_{j,k=0}^{m-1}
\notag\\
&=\prod_{n=0}^{m-1} \frac{h_n^{(\sigma)}}{[\kappa_n^{(\sigma)}]^2},
\end{align}
where $\kappa^{(\sigma)}_n$ is the leading coefficient
of the polynomial $p_n^{(\sigma)}(x^2)$. Thus, Continuous Dual Hahn
polynomials can be related to the Hankel determinants
$D_m^{(\sigma)}$ by properly specializing the parameters
$(a,b,c)$ in each of the two cases, $\sigma=0$ and $\sigma=1$.

Comparing Eqns.~\eqref{mu0-mu1}
and \eqref{mu-three} with the orthogonality weight \eqref{ortCDHP},
we are led to specialize the  parameters $(a,b,c)$ to
the values $(0,1/3,2/3)$ for $\sigma=0$  and $(1,1/3,2/3)$ for
$\sigma=1$. Thus, we have
\begin{equation}\label{poly3a}
p^{(0)}_n(x^2)=S_n\left(\frac{x^2}{36};0,\frac{1}{3},\frac{2}{3}\right)
=\left(\frac{1}{3}\right)_n \left(\frac{2}{3}\right)_n
\Fthreetwo{-n}{\mathrm{i}x/6}{-\mathrm{i}x/6}{1/3}{2/3}{1}\,,
\end{equation}
and
\begin{equation}\label{poly3b}
p^{(1)}_n(x^2)=S_n\left(\frac{x^2}{36};1,\frac{1}{3},\frac{2}{3}\right)
=\left(\frac{4}{3}\right)_n \left(\frac{5}{3}\right)_n
\Fthreetwo{-n}{1+\mathrm{i}x/6}{1-\mathrm{i}x/6}{4/3}{5/3}{1}\,.
\end{equation}
For the normalization constant and the leading coefficient  we have
\begin{equation}
h_n^{(0)}=2\frac{n! (3n)!}{3^{3n-1/2}},\qquad
\kappa_n^{(0)}=\left(-\frac{1}{36}\right)^n,
\end{equation}
and
\begin{equation}
h_n^{(1)}=8\frac{n! (3n+2)!}{3^{3n-1/2}},\qquad
\kappa_n^{(1)}=\left(-\frac{1}{36}\right)^n,
\end{equation}
respectively. The $\Gamma$-function triplication formula,
Eqn.~\eqref{triple}
has again been  used to simplify the expressions for the normalization
constants $h_n^{(0)}$ and  $h_n^{(1)}$. Using \eqref{detD} we obtain
\begin{align}
D_m^{(0)}&= 2^{2m^2-m}\,3^{m^2/2} \prod_{k=0}^{m-1} k!\, (3k)!\,,
\notag\\
D_m^{(1)}&= 2^{2m^2+m}\,3^{m^2/2} \prod_{k=0}^{m-1} k!\, (3k+2)!\,.
\end{align}
Substituting these values into \eqref{D=DD},
taking into account the pre-factor of Eqn.~\eqref{Z=detZ},
and cancelling whatever possible,
we obtain the following expression  for  the
partition function  at $\lambda=\pi/2$ and $\eta=\pi/6$,
\begin{align}
Z_{2m}\Big|_{
\begin{subarray}{l}
\lambda=\pi/2\\
\eta=\pi/3
\end{subarray}
}
&=\left(\frac{1}{2}\right)^{4m^2}\,3^{m^2+m}\,
\frac{m!}{(3m)!}\,
\prod_{k=0}^{m-1}
\left[\frac{(3k+2)!}{(m+k)!}\right]^2
\\
Z_{2m+1}\Big|_{
\begin{subarray}{l}
\lambda=\pi/2\\
\eta=\pi/3
\end{subarray}
}
&=\left(\frac{1}{2}\right)^{(2m+1)^2}\,3^{m^2+2m+1/2}\,
\prod_{k=0}^{m-1}
\left[\frac{(3k+2)!}{(m+k+1)!}\right]^2
\end{align}
Recalling  \eqref{AnZn} which now reads
\begin{equation}
A(N;3)
=2^{N^2}\left(\frac{1}{{\sqrt 3}}\right)^N\,
Z_N\Big|_{
\begin{subarray}{l}
\lambda=\pi/2\\
\eta=\pi/3
\end{subarray}
}\;.
\end{equation}
formulae \eqref{An3} for $3$-enumeration of ASMs are readily recovered.

In conclusion, in this Section
we have given explicit formulae
for the partition function of the six-vertex model with DWBC for some
particular values of parameters $\lambda$ and $\eta$.
Analogous, essentially  equivalent, expressions were already known
from the investigations of ASMs $x$-enumeration. It is worth
emphasizing that the approach presented here
allows to recover these results in a very simple and straightforward way.
The keystone of the whole approach is the nice connection with known
classical orthogonal polynomials that will  now be further exploited
to explore the boundary correlation function in application to the
refined $x$-enumerations of ASMs.

\section{The boundary correlator and refined enumerations of
ASMs}\label{refxenum}

\subsection{Preliminaries}

In this Section we shall  show how the   knowledge of the
suitable set of classical orthogonal polynomials associated to each
of  the three considered cases, can be further exploited to
derive explicit answers for the one point boundary correlation function.
As a consequence the results for the refined $x$-enumerations
of ASMs can be obtained; the result for the refined $3$-enumeration
of ASMs is of primary interest since it was not conjectured previously.

In what follows, for the sake of simplicity and clarity, we shall often
ignore the overall normalization of the boundary correlator.
As already discussed at the end of Section \ref{sec.6vm},
the proper normalization
can always be restored trough the use of Eqn.~\eqref{normcond}.

The basic idea is very simple, stemming from the fact,
see Refs.~\cite{S-75,E-81},
that the polynomials associated to a given orthogonality weight $\mu(x)$
can be in turn represented as determinants,
\begin{equation}\label{polydet}
p_{N-1}(x)= \const \det \mathcal{W},
\end{equation}
where $N\times N$ matrix $\mathcal{W}$ differs from $\mathcal{Z}$,
defined by \eqref{moments},
just only in the  elements of the last column,
\begin{equation}\label{Wmatrix}
\mathcal{W}_{jk} =
\begin{cases}
\mathcal{Z}_{jk} &\text{for}\quad k=0,\ldots, N-2,
\\
x^j
 &\text{for}\quad k=N-1
\end{cases}.
\end{equation}
The boundary one point correlator, see Eqn.~\eqref{HNM},
therefore reads
\begin{equation}\label{bulkcorr}
H_N^{(r)}= \const
\left[\frac{\sin(\lambda+\eta)}{\sin(\lambda-\eta)}\right]^r
\left\{
p_{N-1}(\partial_\eps)
\frac{(\sin\eps)^{r-1}[\sin(\eps-2\eta)]^{N-r}}
{[\sin(\eps+\lambda-\eta)]^{N-1}}\right\}
\bigg|_{\eps=0}.
\end{equation}
This representation is completely general, however it can be fruitfully
exploited only when the explicit form of the polynomial entering it
is known, which happens precisely in each of the  three cases
under consideration, as explained in the previous Section.

Indeed, in these three cases, the polynomials of interest, being of
hypergeometric type,
are known to satisfy the following finite
difference equations with respect to their variable, see, e.g.,
Ref.~\cite{KS-98}. Denoting $y(x)=\MP{n}{\alpha}{x}{\phi}$
for Meixner-Pollaczek polynomials one has
\begin{equation}\label{MPeq}
\rme^{\rmi\phi}(\alpha-\rmi x)y(x+\rmi)
+2\rmi[x\cos\phi-(n+\alpha)\sin\phi]y(x)
-\rme^{-\rmi\phi}(\alpha+\rmi x)y(x-\rmi)=0.
\end{equation}
The Continuous Hahn polynomials satisfies
\begin{equation}\label{CHeq}
B(x)y(x+\rmi)-[B(x)+D(x)+n(n+a+b+c+d-1)]y(x)+D(x)y(x-\rmi x)=0
\end{equation}
where $y(x)=p_n(x;a,b,c,d)$ and
\begin{equation}
B(x)=(c-\rmi x)(d-\rmi x),\qquad
D(x)=(a+\rmi x)(b+\rmi x).
\end{equation}
For the Continuous  Dual Hahn polynomials one has
\begin{equation}\label{DCHeq}
B(x)y(x+\rmi)-[B(x)+D(x)+n]y(x)+D(x)y(x-\rmi x)=0
\end{equation}
where $y(x)=S_n(x^2;a,b,c)$ and
\begin{equation}
B(x)=\frac{(a-\rmi x)(b-\rmi x)(c-\rmi x)}{2\rmi x (2\rmi x -1)},\qquad
D(x)=\frac{(a+\rmi x)(b+\rmi x)(c+\rmi x)}{2\rmi x (2\rmi x +1)}.
\end{equation}
The approach we shall apply here to compute the boundary
correlator is based on the fact  that each of these finite-difference
equations for the polynomials
can be directly translated into a recurrence relation for
the boundary correlator which in turn can be solved explicitly.

The derivation of the recurrence relations
for the boundary correlator in the three cases is quite
similar and will be explained below in detail in each case.
The general idea  underlying the procedure to obtain
the recurrence relation is based on the simple relation
$y(\partial_\eps\pm\rmi)=\rme^{\mp\rmi\eps}y(\partial_\eps)
\rme^{\pm\rmi\eps}$ which allows us to derive from each finite difference
equation a relation of the form
\begin{equation}\label{geneq}
y(\partial_\eps) K_\eps f(\eps)\Big|_{\eps=0}=0.
\end{equation}
Here $K_\eps$ is some linear differential operator
whose form is determined by the finite difference equation,
and $f(\eps)$ is a trial function. Under special choice of this
function \eqref{geneq} becomes a linear recurrence relation for the
boundary correlator.

To find this function we shall use also the fact that the formula
\eqref{bulkcorr} can be further rewritten as follows
\begin{equation}\label{bulkcorrbis}
H_N^{(r)}= \const \Bigl\{
p_{N-1}(\partial_\eps)
[g(\eps)]^{N-1} [\omega(\epsilon)]^{r-1}\Bigr\}\Big|_{\eps=0}
\end{equation}
where $\omega(\eps)$ and $g(\eps)$ are given by
\begin{equation}
\omega(\epsilon):=
\frac{\sin(\lambda+\eta)\sin\eps}{\sin(\lambda-\eta)\sin(\eps-2\eta)}\,,
\qquad
g(\epsilon):=
\frac{\sin(\lambda-\eta)\sin(\eps-2\eta)}
{\sin (2\eta)\sin(\eps+\lambda-\eta)}\,,
\end{equation}
and related to each other as
\begin{equation}
g(\eps)=\frac{1}{\omega(\eps)-1}.
\end{equation}
It is clear that
in order to rewrite Eqn.~\eqref{geneq} as a recurrence relation for
$H_N^{(r)}$ the function $f(x)$ should be chosen of the form
\begin{equation}
f(\eps)=[g(\eps)]^{N-1}  [\omega(\epsilon)]^{r-1}\tau(\eps)
\end{equation}
where $\tau(\eps)$ is still arbitrary. Then, rewriting the  operator
$K_\eps$ in terms of the differential operator
\begin{equation}
D_{\eps}:=\omega\partial_{\omega}=
-\frac{\sin\epsilon\sin(\epsilon-2\eta)}{\sin 2\eta}
\,\partial_\epsilon.
\end{equation}
and reexpressing all quantities in terms of the variable
$\omega$, the expression for
$\tau(\eps)$ can be easily chosen in such a way that Eqn.~\eqref{geneq}
turns into a  recurrence relation for the boundary correlation function.
The actual procedure becomes apparent after turning to the examples.

As a last comment here it should be mentioned
that  the recurrence relation for $H_N^{(r)}$
can equivalently  be viewed as an ordinary differential
equation for the generating function
\begin{equation}\label{HNz}
H_N(z):=\sum_{r=1}^N H_N^{(r)} \,z^{r-1},\qquad H_N(1)=1.
\end{equation}
Thus the problem of solving the obtained recurrence relation
can also be regarded as that  of finding a polynomial solution
to the corresponding differential equation. In all cases considered
below, such differential equations  appear to be at most of the second
order.

\subsection{The free fermion line}

We start with reminding that this is the case
when $\eta=\pi/4$, with the spectral parameter free
to vary within the interval $\pi/4<\lambda<3\pi/4$, or, using
$\phi=2\lambda-\pi/2$ we have $0<\phi<\pi$.
The value $\lambda=\pi/2$ (or $\phi=\pi/2$)
corresponds to $2$-enumerated ASMs.

The boundary correlator in this case is given by the formula
\eqref{bulkcorrbis} where
\begin{equation}\label{og2}
\omega(\eps)=-\cot(\phi/2)\tan\eps,
\qquad
g(\eps)=-\frac{\sin(\phi/2)\cos\eps}{\sin(\eps+\phi/2)},
\end{equation}
and $p_{N-1}(x)$ is given by the formula
\eqref{poly2}, i.e., being the particular case of
the Meixner-Pollaczek polynomial.
The finite-difference equation \eqref{MPeq} in this case reads
\begin{multline}
\rme^{\rmi\phi} \left(\frac{1}{2}- \frac{\rmi x}{4}\right)
p_{N-1}(x+4\rmi)
+2 \rmi\left[ \frac{x}{4}\cos\phi -\left(N-\frac{1}{2}\right)
\sin\phi\right] p_{N-1}(x)
\\
-\rme^{-\rmi\phi}\left(\frac{1}{2}+ \frac{\rmi x}{4}\right)
p_{N-1}(x-4\rmi)=0.
\end{multline}
Using $p_{N-1}(\partial_\eps\pm 4\rmi)=\rme^{\mp 4\rmi\eps}
p_{N-1}(\partial_\eps) \rme^{\pm 4\rmi\eps}$ we derive the condition
\begin{equation}\label{bulkcorr2}
p_{N-1}(\partial_\eps)\
K_{\eps} \
[g(\epsilon)]^{N-1} [\omega(\epsilon)]^{r-1} \tau(\epsilon)
\Big|_{\eps=0}=0
\end{equation}
where $K_{\eps}$ is the first order linear differential operator
\begin{equation}
K_{\eps}=
-\frac{\sin\eps\cos\eps}{\sin\phi}\;\partial_\eps\,\sin(2\eps+\phi)+N-1
\end{equation}
and $\tau(\epsilon)$ is an  arbitrary function which will be suitably
chosen below to turn equation \eqref{bulkcorr2} into a
recurrence relation for the boundary correlator.
Our aim now is to explain how this function can be found
and the recurrence relation can be obtained.

First, using the operator
\begin{equation}
D_\eps=\sin\eps\cos\eps\,\partial_\eps=\omega\partial_\omega
\end{equation}
and changing  to the variable $\omega$ using
\begin{equation}
\frac{\sin(2\eps+\phi)}{\sin\phi}
= -\frac{(\omega-1)(\alpha\omega+1)}{\alpha\omega^2+1}\,,
\qquad
\alpha:=\tan^2(\phi/2),
\end{equation}
we find
\begin{equation}
K_{\eps}
=\Big\{D_\eps (\omega-1)(\alpha\omega+1)
+(N-1)(\alpha\omega^2+1)
\Big\}\frac{1}{\alpha\omega^2+1}.
\end{equation}
Next, let us define the operator $\wt K_\eps$ by
\begin{equation}
K_\eps g^{N-1} =g^{N-1} \wt K_\eps
\end{equation}
where $g$ is given by \eqref{og2}. Using
\begin{equation}
g^{-1} (D_\eps g)= -\frac{\omega}{\omega-1}
\end{equation}
we find
\begin{equation}\label{wtK}
\wt K_\eps
= \Big\{
D_\eps (\alpha\omega+1)-(N-1)\Big\}
\frac{\omega-1}{\alpha\omega^2+1}.
\end{equation}
Now the choice of the function $\tau(\eps)$ is evident, since
in terms of the operator $\wt K_\eps$ the relation
\eqref{bulkcorr2} reads
\begin{equation}
p_{N-1}(\partial_\eps)\, g^{N-1}
\wt K_{\eps} \,
\omega^{r-1} \tau
\Big|_{\eps=0}=0
\end{equation}
and therefore if we choose $\tau(\eps)$ to cancel the factor
standing outside the braces in \eqref{wtK} then we immediately obtain
\begin{equation}
p_{N-1}(\partial_\eps)\,
g^{N-1}\, \Big\{D_\eps (\alpha\omega+1)-(N-1)\Big\}\,
\omega^{r-1} \Big|_{\eps=0}=0.
\end{equation}
Finally, reminding that $D_\eps=\omega\partial_\omega$
this last equation directly leads to the recurrence relation
\begin{equation}\label{rec2}
\alpha r H_N^{(r+1)} - (N-r)  H_N^{(r)} =0
\end{equation}
where we have used \eqref{bulkcorrbis}. Note, that actual
choice of the function $\tau(\eps)$ have been governed to have
all coefficients in the braces in \eqref{wtK} to be polynomials
in $\omega$. Note also the `anti-normal' ordering of the
differential operator there. Below we shall proceed in other cases
exactly in the same way.

Solving the recurrence \eqref{rec2} we find
\begin{equation}
H_N^{(r)}=
\binom{N-1}{r-1}
\frac{\left[\tan^2(\phi/2)\right]^{N-r}}
{\left[1+\tan^2(\phi/2)\right]^{N-1}}\,.
\end{equation}
Here the proper normalization is achieved by directly satisfying
condition \eqref{normcond}.

Thus, the result of Ref.~\cite{BPZ-02} is readily recovered.
Moreover,  recalling \eqref{AnrHnr}, and specializing
$\lambda=\pi/2$, that is $\phi=\pi/2$ or $\tan(\phi/2)=1$,
the expression \eqref{Anr2} for the refined
$2$-enumeration, obtained in \cite{MRR-83,EKLP-92a,EKLP-92b,JPS-98},
is immediately reproduced.

\subsection{The ice point and the refined enumeration of ASMs}

Now we turn to the interesting case of the ice point, which
corresponds to the values $\lambda=\pi/2$ and $\eta=\pi/6$.
In this case the boundary correlator
$H_N^{(r)}$ is equivalent to the refined $1$-enumeration
of ASMs.

In this case the boundary correlator is given by  formula
\eqref{bulkcorrbis} where
\begin{equation}\label{og1}
\omega(\epsilon)=\frac{\sin\eps}{\sin(\eps-\pi/3)},\qquad
g(\epsilon)=\frac{\sin(\eps-\pi/3)}{\sin(\eps+\pi/3)},
\end{equation}
and $p_{N-1}(x)$ is Continuous Hahn polynomial, see
\eqref{poly1}. The finite difference equation \eqref{CHeq} reads
\begin{multline}
\left(\frac{1}{3}-\frac{\rmi x}{6}\right)
\left(\frac{2}{3}-\frac{\rmi x}{6}\right) p_{N-1}(x+6\rmi)
+\left[\frac{x^2}{18}-\frac{4}{9}- N(N-1)\right] p_{N-1}(x)
\\
+\left(\frac{1}{3}+\frac{\rmi x}{6}\right)
\left(\frac{2}{3}+\frac{\rmi x}{6}\right)
p_{N-1}(x-6\rmi)
=0
\end{multline}
Similarly to the previous case,
employing $p_{N-1}(\partial_\eps\pm 6\rmi)=\rme^{\mp 6\rmi\eps}
p_{N-1}(\partial_\eps) \rme^{\pm 6\rmi\eps}$ we obtain
\begin{equation}\label{bulkcorr1}
p_{N-1}(\partial_\eps)\
K_{\eps} \
[g(\epsilon)]^{N-1} [\omega(\epsilon)]^{r-1} \tau(\epsilon)
\Big|_{\eps=0}=0
\end{equation}
where $K_{\eps}$ is the
second order differential operator
\begin{equation}
K_\eps=\frac{1}{9}\sin3\eps\,\partial_\eps^2\,\sin3\eps
+\frac{1}{9}\sin^2 3\eps- N(N-1).
\end{equation}
Now our aim is to obtain the recurrence relation for $H_N^{(r)}$
and find its solution.

The operator $D_\eps$ in this case reads
\begin{equation}
D_\eps=:\omega\partial_\omega
=-\frac{2}{\sqrt 3}\sin\eps\sin(\eps-\pi/3)\partial_\eps.
\end{equation}
Taking into account the identity
\begin{equation}
\sin 3\eps = -4\sin\eps\sin(\eps+\pi/3)\sin(\eps-\pi/3)
\end{equation}
and
\begin{equation}
\sin\eps=-\frac{\sqrt 3}{2}\frac{\omega}{\sqrt{\omega^2-\omega+1}},
\qquad
\sin(\eps+\pi/3)
=-\frac{\sqrt 3}{2}\frac{\omega-1}{\sqrt{\omega^2-\omega+1}},
\end{equation}
we reexpress the operator
$K_{\eps}$ in terms of $\omega$ and the operator $D_\eps$,
with the latter acting from  the very left,
\begin{equation}
K_{\eps}=\Bigl[
D_\eps^2(\omega-1)^2-
D_\eps(\omega^2-1)-N(N-1)(\omega^2-\omega+1)\Bigr]
\frac{1}{\omega^2-\omega+1}.
\end{equation}
Taking into account that
\begin{equation}
g^{-1}(D_\eps g)=-\frac{\omega}{\omega-1},
\end{equation}
for the operator $\wt K_\eps:= g^{-N+1} K_\eps g^{N-1}$ we obtain
\begin{equation}
\wt K_{\eps}  = \Bigl\{
D_\eps^2 (\omega-1)
-D_\eps [(2N-1) \omega+1]+N(N-1)
\Bigr\}
\frac{\omega-1}{\omega^2-\omega+1}.
\end{equation}
Choosing $\tau(\eps)=(\omega^2-\omega+1)/(\omega-1)$
in \eqref{bulkcorr1} allows us to write
\begin{equation}
p_{N-1}(\partial_\eps)\, g^{N-1}
\Bigl\{
D_\eps^2 (\omega-1)
-D_\eps [(2 N -1)\omega+1]+N(N-1)
\Bigr\}\, \omega^{r-1} \Big|_{\eps=0}=0
\end{equation}
which, recalling that $D_{\eps}=\omega\partial_{\omega}$, and
Eqn.~\eqref{bulkcorrbis}, immediately gives us the following
recurrence relation
\begin{equation}\label{recur1}
r(r-2N+1) H_N^{(r+1)}
-(r-N)(N+r-1)H_N^{(r)}=0.
\end{equation}
This recurrence  can be easily solved modulo a
normalization constant
\begin{equation}
H_N^{(r)}= \const  \frac{(N+r-2)!\,(2N-1-r)!}{(r-1)!\,(N-r)!}.
\end{equation}

A possible way to fit the normalization condition \eqref{normcond},
is to consider the generating function
$H_N(z)$ defined via Eqn.~\eqref{HNz}. The result reads
\begin{align}\label{Hnz1}
H_N(z)= \frac{(2N-1)!\,(2N-2)!}{(N-1)!\,(3N-2)!}\;\F{1-N}{N}{2-2N}{z}
\end{align}
where the proper normalization is easily determined
through Chu-Vandermonde identity
\begin{equation}\label{Gauss-sum}
\F{-m}{b}{c}{1}
=\frac{(c-b)_m}{(c)_m};\qquad
(a)_m:=a(a+1)\cdots(a+m-1).
\end{equation}\
Inspecting the coefficient of $z^{r-1}$ in \eqref{Hnz1}
we finally obtain
\begin{equation}\label{Hnr1}
H_N^{(r)}=\frac{\binom{N+r-2}{N-1}\binom{2N-1-r}{N-1}}{\binom{3N-2}{N-1}}.
\end{equation}
The refined $1$-enumeration of ASMs, Eqn.~\eqref{Anr},
immediately follows from the last formula and the
relation \eqref{AnrHnr}.

It is worth to noting that the proof presented here for the refined
$1$-enumeration of ASMs is considerably simpler
in comparison to that of Refs.~\cite{Z-96b,S-02}, which were
based on the inhomogeneous
square ice  partition function formula of Ref.~\cite{I-87}.

\subsection{The $\Delta=-1/2$ symmetric point and
the refined $3$-enumeration of ASMs}\label{ref3enum}

We shall now apply the same approach to compute
the boundary correlator  $H_N^{(r)}$ in the case of
the $\Delta=-1/2$ symmetric point,
i.e., when $\eta=\pi/3$ and $\lambda=\pi/2$. This  case
corresponds  to the refined $3$-enumeration
of ASMs.

As shown in Section \ref{sec.3en}, in this case there are
two sets of Continuous  Dual Hahn polynomials,
with differently specified parameters, see Eqns.~\eqref{poly3a}
and \eqref{poly3b}, which are related to the determinant of the
Hankel matrix $\mathcal{Z}$, see Eqn.~\eqref{Zmat}.
The appearance of two sets of polynomials is due to
the factorization of the Hankel determinant $\det \mathcal{Z}$.
The explicit form of this factorization in turn depends on whether $N$
is odd or even, see Eqn.~\eqref{D=DD}. Such factorization occurs also
for the determinant of matrix $\mathcal{W}$, defined by
\eqref{Wmatrix}. Denoting $D_N(x)=\det \mathcal{W}$, similarly
to \eqref{D=DD} we have
\begin{equation}
D_{2m}^{}(x)=D^{(0)}_{m}\, x\,  D^{(1)}_{m}(x^2),\qquad
D_{2m+1}^{}(x)=D^{(0)}_{m+1}(x^2)\, D^{(1)}_{m},
\end{equation}
where $D_m^{(\sigma)}(x^2)$ stands for the determinants of the matrices
built from the even moments of measures $\mu^{(\sigma)}$, with entries
of the last column replaced by $x^{2j}$. These determinants are
precisely the  Continuous  Dual Hahn polynomials, $
p_{m-1}^{(\sigma)}(x^2)$,
specified by Eqns.~\eqref{poly3a}
and \eqref{poly3b}. Therefore up to overall constants determined
by $D_m^{(1)}$ and $D_m^{(0)}$, we have
\begin{equation}
p_{N-1}(x)=
\begin{cases}
\const p_{m}^{(0)}(x^2) \quad &\text{if}\ N\ \text{is odd}, N=2m+1
\\
\const x\, p_{m}^{(1)}(x^2)\quad &\text{if}\ N\ \text{is even}, N=2m+2
\end{cases}.
\end{equation}
In this way the polynomials appearing in \eqref{bulkcorr} are
expressed in terms  of Continuous Dual Hahn polynomials.

Hence, in the considered case
representation \eqref{bulkcorrbis} acquires the form
\begin{align}
\label{correven}
&\left.H^{(r)}_{2m+2}=
\const p^{(1)}_m(\partial_\eps^2) \,\partial_\eps^{}
\,g^{2m+1}
\omega^{r-1}\right\vert_{\eps=0} \\
\label{corrodd}
&\left.H^{(r)}_{2m+3}= \const p^{(0)}_{m+1}(\partial_\eps^2) \, g^{2m+2}
\omega^{r-1}\right\vert_{\eps=0}
\end{align}
where
\begin{equation}
\omega=\omega(\eps)=-\frac{\sin\eps}{\sin(\eps+\pi/3)}, \qquad
g=\frac{1}{\omega-1}.
\end{equation}
Here we have shifted $m\to m+1$ for the $N$ odd case for later
convenience.

Starting from expressions \eqref{correven} and \eqref{corrodd},
one can derive the recurrence relation for $H_N^{(r)}$ in the  cases
of $N$ even and odd, respectively. Here instead we shall proceed
differently  using the fact that it is possible  to express
the boundary correlator in terms of a single set of polynomials,
and thus  treat  both cases in a unified and simplified way.
Indeed, denoting
\begin{equation}
u_{2m}^{(\sigma)}(x):= p_{m}^{(\sigma)}(x^2),\qquad \sigma=0,1,
\end{equation}
let us consider the polynomials
\begin{equation}\label{poly3tilde}
\tilde u_{2m}(x)=
S_m\left(\frac{x^2}{36};\frac{1}{2},-\frac{1}{6},\frac{1}{6}\right).
\end{equation}
These polynomials arise when one studies the action of the
`forward shift operator' for the  Continuous Dual Hahn polynomials
(see, e.g., Ref.~\cite{KS-98}, Eqn. (1.3.7)),
\begin{equation}\label{u-u}
\tilde u_{2m+2}(x+3\rmi)-\tilde u_{2m+2}(x-3\rmi)=
-\rmi (m+1) \frac{x}{3}\, u^{(1)}_{2m}(x).
\end{equation}
Surprisingly enough, changing the sign
in LHS of this relation gives us again Continuous Dual Hahn polynomials,
which are exactly those defined above as $u_{2n}^{(0)}(x)$, i.e.,
\begin{equation}\label{u+u}
\tilde u_{2m}(x+3\rmi) +\tilde u_{2m}(x-3\rmi)
=2 u^{(0)}_{2m}(x).
\end{equation}
This relation  can be easily proven
by expressing all polynomials on both sides
as truncated hypergeometric series. It is worth to note
that this relation is not a specialization of some general
relation for the Continuous Dual  Hahn polynomials, but is instead
specific for the particular choice of parameters of the polynomials.

As a direct consequence of  relations
\eqref{u-u} and \eqref{u+u} we can rewrite the correlator as
\begin{align} \label{correven2}
H^{(r)}_{2m+2}&= \const \tilde u_{2m+2}(\partial_\eps)
\,\sin 3\eps\, g^{2m+1}
\omega^{r-1}\Big|_{\eps=0},
\\ \label{corrodd2}
H^{(r)}_{2m+3}&=\const  \tilde u_{2m+2}(\partial_\eps)
\,\cos 3\eps\, g^{2m+2}
\omega^{r-1}\Big|_{\eps=0}.
\end{align}
Noticing that
\begin{equation}
\sin 3\eps=-\frac{3\sqrt 3}{2}
\frac{\omega(\omega+1)}{(\omega^2+\omega+1)^{3/2}},\qquad
\cos 3\eps=-
\frac{(\omega-1)(2\omega+1)(\omega+2)}{2(\omega^2+\omega+1)^{3/2}}
\end{equation}
and recalling that $g=1/(\omega-1)$ it can be easily seen
that the correlator possesses the structure
\begin{align} \label{correven3}
H^{(r)}_{2m+2} &= \frac{B_{2m}^{(r-1)}+B_{2m}^{(r-2)}}{2},
\\ \label{corrodd3}
H^{(r)}_{2m+3} &
=\frac{2 B_{2m}^{(r-1)}+ 5 B_{2m}^{(r-2)} +2 B_{2m}^{(r-2)}}{9},
\end{align}
where the quantities $B_{2m}^{(r)}$ are defined as
\begin{equation}\label{B2mr}
B_{2m}^{(r)} := b_{2m}
\tilde u_{2m+2} (\partial_\eps)
\frac{g^{2m+1} \omega^{r+2}}{(\omega^2+\omega+1)^{3/2}}
\bigg|_{\eps=0};\qquad  r=0,1,\dots,2m.
\end{equation}
Here, $b_{2m}$ is some normalization constant; we assume that
\begin{equation}\label{normBnr}
\sum_{r=0}^{2m} B_{2m}^{(r)} =1.
\end{equation}
This condition, together with  Eqns.~\eqref{correven3} and
\eqref{corrodd3}, ensures  that normalization condition
\eqref{normcond} is satisfied.
As in previous cases, the proper normalization will be restored
at the end of computation, according to condition \eqref{normBnr}.

Thus, instead of studying the correlator for
$N$ even and odd separately it is enough to consider the
quantity $B_{2m}^{(r)}$ which is  defined by an essentially similar
formula, Eqn.~\eqref{B2mr}. The procedure developed previously
will be applied now to derive a recurrence relation for $B_{2m}^{(r)}$.
The finite-difference equation \eqref{DCHeq} reads
\begin{equation}
(1+x^2) \Big[\tilde u_{2m+2}(x+6\rmi)-\tilde u_{2m+2}(x-6\rmi)\Big]
-24 \rmi (m+1) x \tilde u_{2m+2}(x)=0.
\end{equation}
Using this equation, like in the previous cases, we can write
\begin{equation}\label{diffop3}
\tilde u_{2m+2}(\partial_\eps)
K_{\eps} f(\eps)
\Big|_{\eps=0}=0
\end{equation}
where $K_{\eps}$ is the second order differential operator
\begin{equation}
K_{\eps}=\frac{1}{\sqrt{3}}
\Big[\sin 3\eps \cos 3\eps\,
\big(\partial_\eps^2+1\big)-6(m+1) \partial_\eps\Big]
\end{equation}
while $f(\eps)$ is some arbitrary function.
Obviously, to obtain a recurrence relation for the quantities
$B_{2m}^{(r)}$, the function $f(\eps)$ is to be chosen of the form
\begin{equation}
f=\tau\, \frac{g^{2m+1}\omega^{r+2}}{(1+\omega+\omega^2)^{3/2}}
\end{equation}
where again $\tau$ is some function to be chosen later.

Taking into account that the differential operator
$D_\eps$ in the considered case reads
\begin{equation}
D_\eps=\omega\partial_{\omega}=
\frac{2}{\sqrt 3}\sin\eps\sin(\eps+\pi/3)
\partial_\eps
\end{equation}
and  passing to the variable $\omega$, we obtain
\begin{multline}
K_{\eps}=D_\eps^2
\frac{(2\omega^2+5\omega+2)(\omega^2-1)}{\omega(\omega^2+\omega+1)}
\\
+D_\eps\left[
\frac{(\omega^2+4\omega+1) (2\omega^2+5\omega+2) (\omega-1)^2}
{\omega(\omega^2+\omega+1)^2}+4 m\frac{(\omega^2+\omega+1)}{\omega}
\right]
\\
-\left[
\frac{(8\omega^4+4\omega^3+57\omega^2+4\omega+8)
(2\omega^2+5\omega+2) (\omega^2-1)}
{4\omega(\omega^2+\omega+1)^3}+4m\frac{\omega^2-1}{\omega}\right].
\end{multline}
Defining  operator $\wt K_\eps$ by the formula
\begin{equation}
K_\eps\, \frac{g^{2m+1}}{(\omega^2+\omega+1)^{3/2}}
= \frac{g^{2m+1}}{(\omega^2+\omega+1)^{3/2}}\, \wt K_\eps
\end{equation}
we may rewrite condition \eqref{diffop3}
\begin{equation}
\tilde u_{2m+2}(\partial_\eps)
\frac{g^{2m+1}}{(\omega^2+\omega+1)^{3/2}}\,
\wt K_{\eps} \tau \omega^{r+2}
\Big|_{\eps=0}=0,
\end{equation}
where
\begin{multline}
\wt K_{\eps}=
\bigg\{
D_\eps^2\,(2\omega^2+5\omega+2)(\omega^2-1)
\\
-D_\eps\Big[
(2\omega^2+5\omega+2)(7\omega^2-2\omega-1)
+4m(\omega^4+5\omega^3+4\omega^2-1)
\Big]
\\
+2(2\omega^2+5\omega+2)(5\omega^2-2\omega+1)
+2m (4\omega^4+ 26\omega^3+17\omega^2+5\omega+2)
\\
+4 m^2 \omega(3\omega^2+4\omega+2)
\bigg\}
\frac{1}{\omega (\omega^2+\omega+1)}.
\end{multline}
Choosing $\tau=\omega+1+\omega^{-1}$
we therefore obtain
\begin{multline}
\tilde u_{2m+2}(\partial_\eps) \frac{g^{2m+1}}{(\omega^2+\omega+1)^{3/2}}
\bigg\{
D_\eps^2\,(2\omega^4+5\omega^3-5\omega-2)
\\
-D_\eps\, \Big[
(14\omega^4+ 31\omega^3+2\omega^2-9\omega-2)+
4m (\omega^4+ 5\omega^3+4\omega^2-1)
\Big]
\\
+
2(10\omega^4+ 21\omega^3+2\omega^2+\omega+2)
+2m (4\omega^4+ 26\omega^3+17\omega^2+5\omega+2)
\\
+4m^2 (3\omega^3+4\omega^2+2\omega)
\bigg\}\omega^{r} \bigg|_{\eps=0}=0.
\end{multline}
This equation leads immediately to the recurrence relation
for  $B_{2m}^{(r)}$. The latter enjoy the symmetry
$B_{2m}^{(r)}=B_{2m}^{(2m-r)}$ which
obviously guarantees  $H_N^{(r)}=H_N^{(N-r+1)}$
(this is also known as the top-bottom or left-right
symmetry of the set of ASMs). To make this, symmetry more apparent
it is convenient to write this recurrence relation in terms of
$E_{m}^{(r)}:=B_{2m}^{(m+r)}$, so that
$E_{m}^{(r)}=E_{m}^{(-r)}$, with $r=-m,\dots,m$.
The recurrence relation reads
\begin{multline}\label{recur3}
2(r-m-2)(r+m+1)\, E_m^{(r-2)}
+(5r^2+10rm+r-3m^2-9m-6)\,
E_m^{(r-1)}
\\
+2(1+8m)r\, E_m^{(r)}
-(5r^2-10rm-r-3m^2-9m-6)\,
E_m^{(r+1)}
\\
-2(r-m-1)(r+m+2)\, E_m^{(r+2)}=0.
\end{multline}
Using this relation one can find recursively
all $E_m^{(r)}$'s for any given $m$ assuming that
$E_m^{(r)}=0$ if $|r|>m$.  However, since the recurrence
relation is five-term it can hardly be solved, e.g., by guessing
its solution. The remaining of this Section is an exposition
of a possible way to solve it explicitly.
As we shall show now this can be done by successive transformations
of the generating function for $B_{2m}^{(r)}$.

Consider the generating function
\begin{equation}\label{Emz}
E_{m}(z):=\sum_{r=-m}^{m} E_{m}^{(r)}\,z^{r}=
z^{-m}\sum_{r=0}^{2m}B_{2m}^{(r)}\, z^r,\qquad
E_{m}(z) = E_{m}(z^{-1}).
\end{equation}
which satisfies, as a direct consequence of recurrence relation
\eqref{recur3}, the following homogeneous second
order linear differential equation:
\begin{multline}\label{diffE}
\bigg\{(z-z^{-1})(2z+5+2z^{-1})(z\partial_z)^2
\\
+\Big[(2z+5+2z^{-1})(3z-2+3z^{-1})+2m(5z+8+5z^{-1})\Big]
z\partial_z
\\
-(z-z^{-1})
\Big[m(6z-1+6z^{-1})+m^2(2z+3+2z^{-1})\Big]
\bigg\}  E_m(z)=0\;.
\end{multline}
The solution of this equation
we are interested in is a polynomial of degree $2m$
times factor $z^{-m}$, see Eqn. \eqref{Emz}.
Let us now consider the substitution
\begin{equation}
z=-\frac{x-q}{qx-1},\qquad  q=\exp(i\pi/3)
\end{equation}
which maps the six singularities of this
equation lying on the real axis of the complex $z$-plane, at points
$z=-1,0,1/2,1,2,\infty$, onto the six roots
of equation $x^6=1$, i.e., $x=1,q,q^2,q^3,q^4,q^5$, lying on the unit
circle of the complex $x$-plane. Note moreover the trivial but useful
identity $1+q^2=q$.
Taking into account symmetry \eqref{Emz} enjoyed by $E_m(z)$,
and the identity $(qx-1)(x-q)=(q x)(x-1+x^{-1})$,
it is easily seen that the function
\begin{equation}\label{Vmx}
V_m(x):=(x-1+x^{-1})^m \,E_m \left(-\frac{x-q}{qx-1}\right)
\end{equation}
is again of the form $x^{-m}$ times a polynomial of order $2m$ in $x$,
symmetric under $x\to 1/x$, i.e., the function
$V_m(\rme^{\rmi\varphi})$ is an even trigonometric polynomial of degree
$m$ in $\varphi$. Equation \eqref{diffE}
translates into the following equation:
\begin{multline}
\bigg\{(x^3-x^{-3})(x\partial_x)^2
\\
+\Big[(x+1+x^{-1})(x^2-5x+12-5x^{-1}+x^{-2})-2m(x+x^{-1})(x^2-5+x^{-2})
\Big]
x\partial_x\\
-(x-x^{-1})\Big[m(x^2-4x+13-4x^{-1}+x^{-2})-m^2(x^2-7+x^{-2})\Big]
\bigg\}
V_m(x)=0.
\end{multline}
At the first glance there is no advantage in considering the
function $V_m(x)$ since the underlying recurrence relation for expansion
coefficients of $V_m(x)$ is even worse than that for
those of $E_m(z)$: it is a seven-term relation. However, if one considers
instead the function,
\begin{equation}\label{hmx}
h_m(x)=(x-x^{-1})^{2m+1} \,(x+2+x^{-1}) V_m(x)
\end{equation}
then it appears that the differential equation
satisfied by $h_m(x)$ contains only integer powers of $x^3$ in their
coefficients
\begin{equation}\label{diffeqFmx}
\bigg\{
(x^3-x^{-3})(x\partial_x)^2
-\Big[3(2m+1)(x^3+x^{-3})-6\Big]x\partial_x
+(3m+1)(3m+2)(x^3-x^{-3})
\bigg\}h_m(x)=0,
\end{equation}
or, using the variable $\phi$, related to $x$ as $x=\exp(\rmi\varphi)$,
we have
\begin{equation}\label{diffeqFmphi}
\biggl\{\partial_\varphi^2
-3\biggl[(2m+1)\cot3\varphi
-\frac{1}{\sin3\varphi}\biggr]\partial_\varphi
-(3m+1)(3m+2)\biggr\}h_m(\mathrm{e}^{\mathrm{i}\varphi})=0.
\end{equation}
It is easy to see that the underlying recurrence relation
now is just a three-term and it appears to be solvable
explicitly.

The particular solution of Eqn.~\eqref{diffeqFmphi}
which we are looking for is the
function $h_m(\mathrm{e}^{\mathrm{i}\varphi})$ being
an antisymmetric trigonometric polynomial
of degree $3m+2$ in $\varphi$, see Eqn.~\eqref{hmx}.
Hence, we are led to use the substitution of the form
\begin{equation}
h(\rme^{\rmi\varphi})=
\sum_{k=0}^{2m+1} \gamma_k \sin[(3m+2-3k)\varphi].
\end{equation}
The coefficients $\gamma_k$ have to satisfy
the recurrence relation
\begin{equation}
(3k+2)(3k+3)\gamma_{k+1}+(3m+2-3k)\gamma_{k}
- \big[3(2m-k)+7\big]\big[3(2m-k)+6\big]\gamma_{k-1}=0.
\end{equation}
Inspecting the explicit form of the first few $\gamma_k$'s,
allows us to guess that the solution of the recurrence relation
is
\begin{equation}
\gamma_{2l}= \gamma_0\,(-1)^l \,\frac{(-m-2/3)_{l}}{(1/3)_{l}}
\binom{m}{l},\qquad
\gamma_{2l+1}= \gamma_0\, (-1)^l\,
\frac{(-m-2/3)_{l+1}}{(1/3)_{l+1}}
\binom{m}{l}\ ,
\end{equation}
that can be easily verified directly.
Here $(a)_l$ stands for Pochhammer symbol, defined previously in
Eqn.~\eqref{Gauss-sum}. Thus, we have obtained
for the function $h_m(x)$ the following expression:
\begin{equation}\label{hm-fg}
h_m(x) = c_m \biggl(g_m(x) + \frac{3m+2}{3m+1}\, f_m(x) \biggr)
\end{equation}
where the functions $f_m(x)$ and $g_m(x)$ are given by
\begin{align}\label{gmx}
g_m(x)
&:= \sum_{k=0}^{m} \binom{m+2/3}{k}\binom{m-2/3}{m-k}
\bigl(x^{3m+2-6k}-x^{-3m-2+6k}\bigr)
\\ \label{fmx}
f_m(x)
&:= \sum_{k=0}^{m} \binom{m+1/3}{k}\binom{m-1/3}{m-k}
\bigl(x^{3m+1-6k}-x^{-3m-1+6k}\bigr)
\end{align}
and $c_m$ is some constant such that $V_m(1)=E_m(1)=1$; this
normalization follows from the normalization  condition
\eqref{normBnr} for $B_{2m}^{(r)}$.

It is worth noticing  that function $h_m(x)$ as given in
Eqns. \eqref{hm-fg}, \eqref{gmx}, \eqref{fmx},
in connection with the problem of refined
$3$-enumeration has also been found by Stroganov
in Ref.~\cite{S-03}, within  a different approach,
using a certain functional equation satisfied by the inhomogeneous
square ice partition function. This functional equation had already been
investigated in \cite{S-01} as the Baxter T-Q equation for the
ground state of XXZ Heisenberg spin-$1/2$ chain at $\Delta=-1/2$
and with odd number of sites $N=2m+1$.

The problem we are facing now is to reconstruct, from the explicit
knowledge of
function $h_m(x)$, the generating function $E_{m}(t)$. This amounts
essentially to divide out the factors $(x-x^{-1})^{2m+1}$ and
$(x+2+x^{-1})$ to find function $V_m(x)$ first, and next
to change back to the original variable $z=(x-q)/(1-qx)$
to recover $E_m(z)$. We shall follow the line
of our previous paper \cite{CP-04} where this procedure
was fulfilled.

To undertake the first step in this program it is useful to note
that the functions $f_m(x)$ and $g_m(x)$
can also be  written as follows
\begin{align}\label{gFF}
g_m(x)&=\frac{\Gamma(m+1/3)}{m!\,\Gamma(1/3)}
\biggl[
x^{3m+2}\F{-m}{-m-2/3}{1/3}{x^{-6}}
\notag\\ & \mspace{212mu}
- x^{-3m-2}\F{-m}{-m-2/3}{1/3}{x^{6}}\biggr],
\\
\label{fFF}
f_m(x)&=\frac{\Gamma(m+4/3)}{m!\,\Gamma(4/3)}
\biggl[
x^{-3m+1}\F{-m}{-m+1/3}{4/3}{x^{6}}
\notag\\  & \mspace{212mu}
- x^{3m-1}\F{-m}{-m+1/3}{4/3}{x^{-6}}
\biggr].
\end{align}
Since the parameters of the hypergeometric functions
entering these expressions
differ by integers one can expect
that $g_m(x)$ and $f_m(x)$ are connected by some three-term
relations via Gauss relations (see, e.g., \S 2.8 of \cite{E-81}).
Indeed, using Gauss relations it can be shown that
\begin{multline}
\F{-m}{-m-2/3}{1/3}{\zeta }
=\frac{3m+4}{2}\;\F{-m-1}{-m-2/3}{4/3}{\zeta }
\\
-\frac{3m+2}{2}\;(1+\zeta )\F{-m}{-m+1/3}{4/3}{\zeta }
\end{multline}
and therefore one can express the function
$g_m(x)$ in terms of $f_m(x)$ and $f_{m+1}(x)$:
\begin{equation}\label{gff}
g_m(x)=\frac{3m+2}{2(3m+1)}\;
(x^3+x^{-3})\,f_m(x) -
\frac{3(m+1)}{2(3m+1)}\; f_{m+1}(x).
\end{equation}
Substituting \eqref{gff} into \eqref{hm-fg} we obtain an analogous
formula for function $h_m(x)$:
\begin{equation}\label{hff}
h_m(x)=c_m\,\frac{3m+2}{2(3m+1)}
\left[
(x^3+2+x^{-3})\, f_m(x) - \frac{3m+3}{3m+2}\; f_{m+1}(x)
\right].
\end{equation}
Introducing now the function $Q_m(x)$, implicitly defined by
\begin{equation}\label{fQ}
f_m(x)=(x-x^{-1})^{2m+1} Q_m(x)
\end{equation}
the factor $(x-x^{-1})^{2m+1}(x+2+x^{-1})$ can be formally
extracted in expression \eqref{hff} for $h_m(x)$, thus
giving us a representation for $V_m(x)$ in terms of
$Q_m(x)$ and $Q_{m+1}(x)$,
\begin{equation}\label{VQQ}
V_m(x)= c_m\,\frac{3m+2}{2(3m+1)}
\left[(x-1+x^{-1})^2\,Q_m(x)- \frac{3m+3}{3m+2}\,
(x-2+x^{-1})\, Q_{m+1}(x)
\right].
\end{equation}

The meaning of this procedure becomes apparent by noticing that
the function $Q_m(x)$ can be found explicitly
from expression \eqref{fFF} for the function $f_m(x)$
in virtue of the so-called cubic transformation for the
Gauss hypergeometric function. The details of this
derivation are given in Appendix A.
For function $Q_m(x)$ the following explicit formula
is valid
\begin{equation}\label{QF}
Q_m(x)
=\frac{(2m)!}{3^m\, (m!)^2}\;
\biggl(\frac{qx^{-1}-q^{-1}x}{q-q^{-1}}\biggr)^m
\F{-m}{m+1}{-2m}{\frac{q x-q^{-1} x^{-1}}{qx^{-1}-q^{-1}x}}.
\end{equation}
Hence, function $V_m(x)$ can be found by inserting
this expression into \eqref{VQQ}, thus giving
\begin{multline}  \label{Vpp}
V_m(x)
=\frac{(2m)!\,(2m+1)!}{m!\, (3m+1)!}
\\ \times
\biggl[(x-1+x^{-1})^2\,
\biggl(\frac{qx^{-1}-q^{-1}x}{q-q^{-1}}\biggr)^m
\F{-m}{m+1}{-2m}{\frac{q x-q^{-1} x^{-1}}{qx^{-1}-q^{-1}x}}
\\
-\frac{2(2m+1)}{3m+2}\, (x-2+x^{-1})\,
\biggl(\frac{qx^{-1}-q^{-1}x}{q-q^{-1}}\biggr)^{m+1}
\\ \times
\F{-m-1}{m+2}{-2m-2}{\frac{q x-q^{-1} x^{-1}}{qx^{-1}-q^{-1}x}}
\biggr].
\end{multline}
Here we have written the expression for $V_m(x)$ taking into account
also the proper normalization of this function,
$V_m(1)=1$. The normalization
can be verified by virtue of Chu-Vandermonde identity
\eqref{Gauss-sum} and it is equivalent to the choice
\begin{equation}
c_m=(3m+1)\,\frac{3^{m+1} m!\,(2m+2)!}{(3m+3)!}\;.
\end{equation}
of this constant in Eqns.~\eqref{hm-fg} and \eqref{VQQ}.
It is worth to mention that functions $Q_m(x)$,  Eqn.~\eqref{QF},
and $V_m(x)$, Eqn.~\eqref{Vpp},
are the first and the second solution, respectively, of
Baxter T-Q equation for the ground state of XXZ Heisenberg
$\Delta=-1/2$ spin chain  with odd number  of sites $N=2m+1$,
see Refs.~\cite{PS-99,S-01,S-02,S-03}.

To obtain function $E_m(z)$ from the given expression
for $V_m(x)$ one can use the formula
\begin{equation}
E_m(z)=3^{-m} (z+1+z^{-1})^m  V_m\left(\frac{z+q}{zq+1}\right)
\end{equation}
which is the inverse of \eqref{Vmx}. Applying
this transformation to \eqref{Vpp} we obtain
\begin{multline}\label{Pmtpre}
E_m(z)=  \frac{(2m)!\,(2m+1)!}{3^m\, m!\,(3m+1)!}\,
\frac{1}{(z+1+z^{-1})^{2}}
\bigg[9(z+2)^{m}\F{-m}{m+1}{-2m}{\frac{z^{-1}+2}{z+2}}
\\
+\frac{2(2m+1)}{3m+2}\,
(z-2+z^{-1})
(z+2)^{m+1}\F{-m-1}{m+2}{-2m-2}{\frac{z^{-1}+2}{z+2}}
\bigg].
\end{multline}
This expression is however  not the  final answer yet, since
the factor  $(z+1+z^{-1})^{2}$  in the denominator is to be cancelled
explicitly (recall that $E_m(z)$ is to be $z^{-m}$ times
a polynomial of degree $2m$ in $z$, hence the expression in the brackets
contains implicitly the proper factor $(z+1+z^{-1})^2$).
This can be achieved using again
Gauss relations; for  details, see  Appendix B.
The final result is
\begin{multline}\label{Efinal}
E_m(z)=
\frac{(2m)!\,(2m+2)!}{3^m\,(m+1)!\,(3m+2)!}
\biggl[(2m+1)(z+2)^m\,
\F{-m}{m+2}{-2m-1}{\frac{1+2z}{z(z+2)}}
\\
-3m\,(z+2)^{m-1} \F{-m+1}{m+2}{-2m}{\frac{1+2z}{z(z+2)}}\biggr].
\end{multline}
This completes the derivation of function $E_m(x)$.

We can now extract from this expression closed formulae for
the coefficients
$B_{2m}^{(r)}$, which are related to $E_m(z)$ through Eqn.~\eqref{Emz}.
To find these coefficients we expand
\eqref{Efinal} in power series in $z$, thus expressing $E_{m}(z)$,
as a triple sum, and next we apply  Chu-Vandermonde formula
\eqref{Gauss-sum} to make the sum with respect to the index
defining the hypergeometric
series in \eqref{Efinal}, thus expressing $E_{m}(z)$ as a double sum.
These two summations can be rearranged in such a way that one of them
becomes with respect to $r$ while the other one defines the
coefficients of power expansion in $z$. We obtain
\begin{multline} \label{balpha}
B_{2m}^{(r)}=\frac{(2m+1)!\,m!}{3^m\,(3m+2)!}
\sum_{\ell=\max(0,r-m)}^{[r/2]}
(2m+2-r+2\ell)\binom{3m+3}{r-2\ell}
\\ \times
\binom{2m+\ell-r+1}{m+1} \binom{m+\ell+1}{m+1}2^{r-2\ell}.
\end{multline}
Here $[r/2]$ denotes integer part of $r/2$.
This  expression for $B_{2m}^{(r)}$ indeed solves the five terms
recurrence relation \eqref{recur3}  (recall that
$E_{m}^{(r)}=B_{2m}^{(m+r)}$).

We would like also to mention that formula \eqref{balpha}
can also be written in terms of terminating hypergeometric series,
for instance, as follows:
\begin{multline}\label{bFF}
B_{2m}^{(r)}=
\frac{2^{r}\binom{3m+3}{r} \binom{2m+1-r}{m+1}}
{3^m\binom{3m+2}{m+1}}
\Biggl[
2 \Ffourthree{-(r-1)/2}{-r/2}
{m+2}{2m+2-r}{(3m+4-r)/2}{(3m+5-r)/2}
{m-r+1}{\frac{1}{4}}
\\
-\frac{r}{m+1}
\Ffourthree{-(r-1)/2}{-r/2+1}
{m+2}{2m+2-r}{(3m+4-r)/2}{(3m+5-r)/2}
{m-r+1}{\frac{1}{4}}
\Biggr].
\end{multline}
This formula is valid for $r=0,1,\ldots,m$
(a similar expression for $r=m+1,m+2,\dots,2m$ can be simply
obtained through the replacement $r\to 2m-r$
in RHS of \eqref{bFF}. The two ${}_4F_3$ in \eqref{bFF} can be further
combined into a single ${}_5F_4\,$. Analogous formulae for $B_{2m}^{(r)}$
in terms of terminating hypergeometric series of argument $4$ may
be written down as well.
Analyzing these expressions, however it seems to be hard
to perform the sum in \eqref{balpha} in a closed form,
even if very suggestive
similarities can be found with known summation formulae, see
\S\S 7.5 and 7.6, especially \S 7.6.4, of Ref.~\cite{PBM-III}.

The complete expression for the boundary correlator
$H_N^{(r)}$ can be readily obtained by inserting \eqref{balpha}
or \eqref{bFF} into \eqref{correven3} and \eqref{corrodd3}.
Finally, the refined $3$-enumeration
of ASMs, $A(N,r;3)$, follows from multiplying the result for
$H_N^{(r)}$ by the total number of $3$-enumerated ASMs, $A(N,3)$,
see \eqref{AnrHnr} and \eqref{An3}.

\section{Conclusion}

The main purpose of the present paper was to point out
the close relation between ASM enumerations and
some classical orthogonal polynomials.
In particular, in  Section \ref{xenume}
we have shown an  alternative way to recover
known results for the partition function of the six-vertex model
with DWBC,  in the three cases
of the $\Delta=0$ line, and the symmetric $\Delta=1/2$ and
$\Delta=-1/2$ points, corresponding to
$2$-, $1$- and $3$-enumeration of ASMs, respectively. The derivation we
have presented in Section \ref{xenume} is in our opinion
extremely simple and straightforward.
It is based on the fact that the Hankel determinant entering the
representation for the partition function can be  naturally
related in these three cases to  Meixner-Pollaczek, Continuous Hahn,
and Continuous Dual Hahn polynomials,  respectively.

It is to emphasize that the three considered cases, $x=1,2,3$, are the
only ones  in which `factorized' answers for ASMs' $x$-enumerations are
known to exist. The fact that no set of polynomials corresponding to
other choices of the parameters can be found in the Askey scheme
strongly suggests that no `factorized' answer exist  for other values
of $x$. However, one might speculate that for some values of the
crossing parameter $\eta$, the partition function of the six-vertex
model with DWBC might still admit a `factorized' form, but in terms of
$q$-numbers (with $q$ related to $\eta$ through $q=\rme^{2\rmi\eta}$),
and  obtained via some suitable  $q$-polynomial, possibly of
Askey-Wilson type. Concerning this, it is worth mentioning a recent
paper \cite{ILR-04} which contains rather promising preliminary
results in this direction. Speculating further,  an interesting
problem which we would like  just to hint here concerns a possible
relation of such $q$-polynomials as $q$ tends to a cubic, quartic or
sixth root of unity, with the three sets of classical polynomials
mentioned above. In this respect, recall that in  paper \cite{Z-96b}
$q$-Legendre polynomials were used while in the present paper
Continuous Hahn polynomials have shown to play an analogous role.

In Section \ref{refxenum} we have used the knowledge of appropriate
orthogonal polynomials for the cases under investigation to derive
recurrence relations for the boundary one point correlator $H_N^{(r)}$.
In the first two cases, such recurrences are trivially solved,
and known result for the closely related problems of
ASMs'refined $1$- and $2$-enumerations are easily reproduced.
The same approach is applied to the symmetric
$\Delta=-1/2$ point, but the resulting recurrence relation for
$H_N^{(r)}$ appears very intricate, not being two-term
(like in cases of $2$- and $1$-enumerations) but rather five-term.
Even though the differential equation  for the corresponding
generating function is not of hypergeometric type, it has been
shown to be solvable in terms
of a suitable linear combination  of hypergeometric functions.
The so-called cubic transformation for the Gauss hypergeometric
function has  been applied to work out   an explicit formula
for the refined $3$-enumeration of ASMs, $A(N,r;3)$. The latter
appears not to be writable  as a single hypergeometric term,
i.e., not to be `round' (or `smooth'), contrarily to other known
expressions for enumerations of ASMs, and it hardly could have been
conjectured on the basis, e.g., of computer experiments.

\section*{Acknowledgments}

We acknowledge financial support from MIUR PRIN programme (SINTESI 2004).
One of us (A.G.P.) was also supported in part by Russian Foundation
for Basic Research, under
RFFI grant No. 04-01-00825, and by the programme
Mathematical Methods in Nonlinear Dynamics of
Russian Academy of Sciences.
This work was been partially
done within the European Community
network EUCLID (HPRN-CT-2002-00325).

\appendix
\renewcommand{\theequation}{A.\arabic{equation}}
\setcounter{equation}{0}
\section*{Appendix A}

We reproduce here for the sake of completeness
the proof of Eqn.~\eqref{QF} given in a previous paper \cite{CP-04}.

The key identity which is to be used here is
the so-called cubic transformation of Gauss hypergeometric
function \cite{E-81} which in its most symmetric form reads:
\begin{multline}\label{cubic}
\frac{\Gamma(a)}{\Gamma(2/3)}\;
\F{a+1/3}{a}{2/3}{\zeta^3} - \omega^{-1} \zeta\,
\frac{\Gamma(a+2/3)}{\Gamma(4/3)}\;
\F{a+1/3}{a+2/3}{4/3}{\zeta^3}
\\
= 3^{-3a+1}{\Bigl(\frac{1-\zeta}{1-\omega}\Bigr)^{-3a}}
\frac{\Gamma(3a)}{\Gamma(2a+2/3)}\;
\F{a+1/3}{3a}{2a+2/3}{\omega\frac{\zeta -\omega}{1-\zeta }}.
\end{multline}
Here $\omega$ is a primitive cubic root of unity,
$\omega=\exp(\pm2\mathrm{i}\pi/3)$, and $a$ is arbitrary parameter.
To show that indeed the cubic transformation is relevant to
our case, let us rewrite \eqref{fFF} in the form
consistent with LHS of \eqref{cubic}.
Taking into account that
\begin{equation}
\F{-m}{-m+1/3}{4/3}{\zeta } =\frac{\Gamma(1/3)\,\Gamma(4/3)}
{\Gamma(-m+1/3)\,\Gamma(m+4/3)} (-\zeta )^m
\F{-m}{-m-1/3}{2/3}{\zeta ^{-1}}
\end{equation}
and
\begin{equation}
\frac{\Gamma(1/3)}{\Gamma(m+4/3)}
=(-1)^{m+1} \frac{\Gamma(-m-1/3)}{\Gamma(2/3)}\;,\qquad
\frac{\Gamma(m+4/3)}{\Gamma(-m+1/3)}
=(-1)^m \frac{(3m+1)!}{3^{3m+1}\,m!}
\end{equation}
it is easy to see that \eqref{fFF} can be rewritten in the form
\begin{multline}\label{f-ready}
f_m(x)=\frac{(-1)^{m+1}(3m+1)!}{3^{3m+1}\, (m!)^2}\;
x^{3m+1}\biggl[
\frac{\Gamma(-m-1/3)}{\Gamma(2/3)}\;
\F{-m}{-m-1/3}{2/3}{x^{-6}}
\\
+x^{-2} \frac{\Gamma(-m+1/3)}{\Gamma(4/3)}\;
\F{-m}{-m+1/3}{4/3}{x^{-6}}
\biggr].
\end{multline}
Clearly, both terms in the brackets are the same as in LHS of
\eqref{cubic} provided the parameter $a$ is specialized to the
value $a=-m-1/3$.

To apply the cubic transformation to \eqref{f-ready}
we first define
\begin{equation}\label{waz-in}
W(a;\zeta ) :=
\frac{\Gamma(a)}{\Gamma(2/3)}\;
\F{a+1/3}{a}{2/3}{\zeta ^3} + \zeta \,
\frac{\Gamma(a+2/3)}{\Gamma(4/3)}\;
\F{a+1/3}{a+2/3}{4/3}{\zeta ^3}
\end{equation}
so that
\begin{equation}\label{fW}
f_m(x)=\frac{(-1)^{m+1}(3m+1)!}{3^{3m+1}\, (m!)^2}\;
x^{3m+1} W(-m-1/3;x^{-2}).
\end{equation}
Next, we note that for a sum of two terms one can always write
\begin{equation}
X+Y =\frac{q}{q-q^{-1}}\;\bigl(X-q^{-2}Y\bigr)-
\frac{q^{-1}}{q-q^{-1}}\bigl(X-q^2Y\bigr)
\end{equation}
and if $q=\exp(\mathrm{i}\pi/3)$, which is exactly the case,
one can set $\omega=q^2$
for the first pair of terms and $\omega=q^{-2}$ for the second one.
This recipe allows one to apply the cubic transformation,
that gives
\begin{multline} \label{waz-out}
W(a;\zeta )=
\frac{3^{-3a+1}\,\Gamma(3a)}{\Gamma(2a+2/3)}\;
\frac{(1-\zeta )^{-3a}}{q(1-q^2)^{-3a+1}}
\biggl[
\F{a+1/3}{3a}{2a+2/3}{q^2\frac{\zeta -q^2}{1-\zeta }}
\\
+q^{3a+1}
\F{a+1/3}{3a}{2a+2/3}{q^{-2}\frac{\zeta -q^{-2}}{1-\zeta }}
\biggr].
\end{multline}
To obtain a new formula for $f_m(x)$ via \eqref{fW} we have
to evaluate now the limit $a\to-m-1/3$ of \eqref{waz-out}.
The limit of the pre-factor can be easily found due to
\begin{equation}
\lim_{a\to-m-1/3}\frac{\Gamma(3a)}{\Gamma(2a+2/3)}
=\frac{2}{3}\;\frac{(-1)^{m+1}(2m)!}{(3m+1)!}\;.
\end{equation}
To find the limit of the expression in the brackets in \eqref{waz-out}
we note that
\begin{equation}
q^2\frac{\zeta -q^2}{1-\zeta }=:s\;\qquad
q^{-2}\frac{\zeta -q^{-2}}{1-\zeta }=1-s
\end{equation}
and hence the following formula can be used
\begin{multline}\label{limfin}
\lim_{a\to-m-1/3}\biggl[
\F{a+1/3}{3a}{2a+2/3}{s}
+q^{3a+1}
\F{a+1/3}{3a}{2a+2/3}{1-s}
\biggr]
\\
=\frac{3}{2}\, \F{-m}{-3m-1}{-2m}{s}.
\end{multline}
Formula \eqref{limfin} can be proved, for instance,
by virtue of standard analytic continuation formulae for the
hypergeometric function
(see, e.g., Eqns.~(1) and (2) in \S 2.10 of \cite{E-81}).
Collecting formulae we arrive to the expression
\begin{equation}
W(-m-1/3;\zeta )=- \frac{3^{2m+1}\,(2m)!}{(3m+1)!}
\frac{(1-\zeta )^{3m+1}}{(q-q^{-1})^{m}}
\F{-m}{-3m-1}{-2m}{q^2\frac{\zeta -q^2}{1-\zeta }}.
\end{equation}
Finally, substituting this expression into \eqref{fW}
and using the identity
\begin{equation}
\F{a}{b}{c}{\zeta }
=(1-\zeta )^{-a} \F{a}{c-b}{c}{\frac{\zeta }{\zeta -1}}
\end{equation}
we obtain
\begin{equation}
f_m(x)=
\frac{(2m)!}{3^m\,(m!)^2}\;
(x-x^{-1})^{2m+1}
\biggl(\frac{qx^{-1}-q^{-1}x}{q-q^{-1}}\biggr)^m
\F{-m}{m+1}{-2m}{\frac{q x-q^{-1} x^{-1}}{qx^{-1}-q^{-1}x}}.
\end{equation}
Obviously, this expression leads directly to \eqref{QF}
which is thus proved.

\renewcommand{\theequation}{B.\arabic{equation}}
\setcounter{equation}{0}
\section*{Appendix B}

Here  we explain how the final expression for function $E_m(z)$,
Eqn.~\eqref{Efinal}, can be obtained from Eqn.~\eqref{Pmtpre}.

We begin with noting that
the hypergeometric functions (polynomials) which enters
expression \eqref{Pmtpre} belong to the class of the functions
\begin{equation}\label{Fab}
\F{a}{b}{a-b+1}{\zeta}
\end{equation}
which are, in the case of $a$ being a negative integer,
possess the symmetry
\begin{equation}
\F{a}{b}{a-b+1}{\zeta}=\zeta^{-a} \F{a}{b}{a-b+1}{\frac{1}{\zeta}}.
\end{equation}
Obviously, since $E_m(z)$ is symmetric with respect to $z\to z^{-1}$,
it is natural to deal only with the functions of the form \eqref{Fab}
when transforming the expression for $E_m(z)$.
It is convenient to use the notation
\begin{equation}
\Psi_m^{(k)} (u)
:=(z+2)^{m}\F{-m}{k+1}{-m-k}{\frac{z^{-1}+2}{z+2}}\;,
\qquad u:=z+1+z^{-1}.
\end{equation}
Note that $\Psi_m^{(k)}(u)$ is a polynomial of degree $m$ in $u$.
Using Gauss relations the following identities can be proven
\begin{align}\label{Pdec}
\Psi_{m+1}^{(k)}(u)&=(u+3)\,\Psi_{m}^{(k)}(u)
-\frac{m(m+2k+1)}{(m+k+1)(m+k)}\,(2u+3)\,\Psi_{m-1}^{(k)}(u)
\\ \label{Pmix}
\Psi_{m}^{(k+1)}(u)&=\frac{m+2k+2}{2(m+k+1)}\,\Psi_m^{(k)}(u)
+\frac{m}{2(m+k+1)}\,(u+3)\,\Psi_{m-1}^{(k+1)}(u)\;.
\end{align}
Returning to \eqref{Pmtpre} we note that
in terms of $\Psi_m^{(k)}$'s the generating function $E_m(z)$
reads
\begin{equation}
E_m(z)=
\frac{(2m)!\,(2m+1)!}{3^m\, m!\,(3m+1)!}\,
\frac{1}{u^2}\left\{9\,\Psi_m^{(m)}(u)+
\frac{2(2m+1)}{3m+2}\,(u-3)\,\Psi_{m+1}^{(m+1)}(u)\right\}.
\end{equation}
Applying \eqref{Pdec}, with $k=m+1$,  gives us
\begin{multline}
E_m(z) =
\frac{(2m)!\,(2m+1)!}{3^m\, m!\,(3m+1)!}\,
\bigg\{
\frac{2(2m+1)}{3m+2}\,\Psi_{m}^{(m+1)}(u)
-\frac{6m}{3m+2}\,\Psi_{m-1}^{(m+1)}(u)
\\
+\frac{9}{u^2}\left[-\frac{2(2m+1)}{3m+2}\,\Psi_{m}^{(m+1)}(u)
+\frac{m}{3m+2}\,(u+3)\,\Psi_{m-1}^{(m+1)}(u)
+\Psi_m^{(m)}(u)\right]\bigg\}.
\end{multline}
Now, the relation \eqref{Pmix} with $k=m$ shows that the expression
in the brackets is zero. Hence we find
\begin{equation}
E_m(z) =
\frac{(2m)!\,(2m+2)!}{3^m\, (m+1)!\,(3m+2)!}\,
\Big[
(2m+1)\,\Psi_{m}^{(m+1)}(u)
-3m\,\Psi_{m-1}^{(m+1)}(u)\Big].
\end{equation}
Restoring the original notations,  we arrive to expression \eqref{Efinal}.


\end{document}